\documentclass{cplslarge}%
\usepackage[authoryear]{natbib}
\usepackage{makeidx}

\usepackage[switch]{lineno}

\usepackage[bookmarks = true, bookmarksnumbered = true, pdfpagemode =None, pdfstartview = FitH, pdfpagelayout = SinglePage, colorlinks = true, urlcolor = red, citecolor = blue]{hyperref}

\usepackage{graphics}
\makeindex

\global\chapterreferencefalse
\usepackage[sectionbib]{chapterbib}

\begin{document}
\frontmatter
%  \maketitle
%\tableofcontents
%  \listoffigures
%  \listoftables
%  \listofcontributors
% \editedlistofcontributors
%  \printphotogallery
%  \include{acknow}
%  \include{foreward}
\mainmatter
%  \part{Getting Started}
%  \include{Introchap}% introduction
%  \include{Newchap}% chapter
  
\author[Fletcher, Sromovsky, Hue, Moses, Guerlet, West, \&\ Koskinen]{l.n. fletcher$^1$, l. sromovsky$^2$, v. hue$^3$, j.i. moses$^4$, s. guerlet$^5$, r.a. west$^6$, and t. koskinen$^7$}
\chapter{Saturn's Seasonal Atmosphere at Northern Summer Solstice}

%\linenumbers
\footnotesize
$^1$ School of Physics and Astronomy, University of Leicester, University Road, Leicester, LE1 7RH, UK.\\
$^2$ Space Science and Engineering Center, University of Wisconsin, Madison, WI 53706, USA. \\
$^3$ Southwest Research Institute, Department of Space Science, Space Science and Engineering Division, 6220 Culebra Road, San Antonio, TX 78238-5166, USA.\\
$^4$ Space Sciences Institute, 4765 Walnut St, Suite B, Boulder, CO 80301, USA.\\
$^5$ Laboratoire de Meteorologie Dynamique / CNRS / Sorbonne Université, 4 Place Jussieu, 75252 Paris, France.\\
$^6$ NASA Jet Propulsion Laboratory, 4800 Oak Grove Drive, Pasadena, CA 91109, USA. \\
$^7$ Lunar and Planetary Laboratory, University of Arizona, 1629 E. University Blvd., Tucson, AZ 85721, USA.

\section*{Copyright Notice}
The Chapter, ``Saturn’s Seasonal Atmosphere at Northern Summer Solstice'', is to be published by Cambridge University Press as part of a multi-volume work edited by Kevin Baines, Michael Flasar, Norbert Krupp, and Thomas Stallard, entitled ``Cassini at Saturn: The Grand Finale" (`the Volume')
 
\copyright  ~in the Chapter, L.N. Fletcher, L. Sromovsky, V. Hue, J.I. Moses, S. Guerlet,  R.A. West and T. Koskinen,

\copyright ~in the Volume, Cambridge University Press
 
NB: The copy of the Chapter, as displayed on this website, is a draft, pre-publication copy only. The final, published version of the Chapter will be available to purchase through Cambridge University Press and other standard distribution channels as part of the wider, edited Volume, once published. This draft copy is made available for personal use only and must not be sold or re-distributed.

\normalsize

\section*{Abstract}
The incredible longevity of Cassini's orbital mission at Saturn has provided the most comprehensive exploration of a seasonal giant planet to date.  This review explores Saturn's changing global temperatures, composition, and aerosol properties between northern spring and summer solstice (2015-2017), extending our previous review of Cassini's remote sensing investigations \citep[2004-2014,][]{18fletcher_book} to the grand finale.  The result is an unprecedented record of Saturn's climate that spans almost half a Saturnian year, which can be used to test the seasonal predictions of radiative climate models, neutral and ion photochemistry models, and atmospheric circulation models.  Hemispheric asymmetries in tropospheric and stratospheric temperatures were observed to reverse from northern winter to northern summer.   Spatial distributions of hydrocarbons and para-hydrogen shifted in response to atmospheric dynamics (e.g., seasonally-reversing circulations, polar stratospheric vortex formation, equatorial stratospheric oscillations, and inter-hemispheric transport).  Upper tropospheric and stratospheric aerosols exhibited changes in optical thickness that modulated Saturn's visible colours (from blue hues to a golden appearance in the north near solstice), reflectivity, and near-infrared emission.  Numerical simulations of radiative balance and photochemistry effectively reproduce the observed seasonal change and phase lags, but discrepancies between models and data still persist, indicating a crucial role for atmospheric dynamics and the need to couple chemical and radiative schemes to the next generation of circulation models.  With Cassini's demise, an extended study of Saturn's seasons, from northern summer to autumn, will require the capabilities of ground- and space-based observatories, as we eagerly await the next orbital explorer at Saturn.

%%%%%%%%%%%%%%%%%%%%%%%%%%%%%%%%%%%%%%%%%%%%%
%%%%%%%%%%%%%%%%%%%%%%%%%%%%%%%%%%%%%%%%%%%%%
%%%%%%%%%%%%%%%%%%%%%%%%%%%%%%%%%%%%%%%%%%%%%
\section{Introduction: Cassini's First Decade}
\label{intro}

Over thirteen years of unprecedented orbital exploration, the Cassini spacecraft provided our first comprehensive characterisation of the seasonal evolution of a Giant Planet atmosphere.  Its remote sensing investigations explored the shifting temperatures, aerosol properties, and spatial distributions of gaseous species, revealing the delicate interplay between atmospheric circulation, photochemistry, radiative heating/cooling, and cloud microphysics.  Given Saturn's $26.7^\circ$ axial tilt and 29.5-year orbit, only a long-lived orbiting spacecraft could monitor these processes over seasonal timescales, providing access to both the sunlit summer hemisphere and the darkness of Saturn's winter.  In \citet{18fletcher_book}, we reviewed our knowledge of the physical and chemical processes shaping the seasonal giant during the first decade of Cassini observations (2004-2014).  The Cassini results were compared to ground-based \citep[e.g.,][]{75gillett, 89gezari, 05orton, 05greathouse}, Pioneer 11 \citep{80orton}, and Voyager \citep{81hanel, 82hanel} infrared observations of thermal emission, and reflected sunlight measurements from Hubble and Earth-based observers \citep[e.g.,][]{92karkoschka, 01stam, 05perez-hoyos, 05karkoschka} from the mid-1970s to the early 21st century.  
\index{Voyager}\index{Pioneer}\index{Hubble}

Cassini arrived at Saturn in July 2004, shortly after northern winter solstice in October 2002 (a heliocentric solar longitude $L_s=270^\circ$), and its prime and extended missions monitored atmospheric changes through northern spring equinox in August 2009 ($L_s=0^\circ$), and into northern mid-spring in 2014 ($L_s\sim55^\circ$).  During this time, the southern summer hemisphere was receding into the darkness of autumn, and the northern winter hemisphere was emerging into sunlight for the first time in 15 years.  In this review, we extend this analysis from 2015 ($L_s\sim60^\circ$) through northern summer solstice in May 2017 ($L_s=90^\circ)$, all the way to the end of Cassini's mission in September 2017 ($L_s=93^\circ$) and beyond, using ground-based observatories.  This extensive seasonal coverage, which encompassed Cassini's proximal orbits and grand finale in 2016-17, means that we have a database of Saturn's climate spanning almost half a Saturnian year, and the period between perihelion ($L_s=280^\circ$, June 2003) and aphelion ($L_s=100^\circ$, April 2018).
\index{Heliocentric Longitude}
\index{solstice}\index{equinox}

The first decade of Cassini observations revealed hemispheric asymmetries in tropospheric and stratospheric temperatures, aerosol opacity, and chemical composition \citet{18fletcher_book}.  These large-scale asymmetries were superimposed onto smaller-scale dynamic perturbations, such as gradients in thermal emission and reflectivity associated with the cool anticyclonic zones (equatorward of prograde jets) and warm cyclonic belts \citep[poleward of prograde jets,][]{83conrath, 06vasavada, 07fletcher_temp, 09delgenio, 20fletcher_beltzone}.  In addition, the hemispheric asymmetries were perturbed by the equatorial enhancements in tropospheric phosphine \citep[a disequilibrium species,][]{09fletcher_ph3}, ammonia \citep[the key condensable forming Saturn's top-most clouds,][]{11fletcher_vims, 13laraia, 13janssen}, and para-hydrogen \citep[a tracer of vertical motion,][]{10fletcher_seasons, 16fletcher}; the response of the stratospheric composition to ring-shadowing \citep{09guerlet, 15sylvestre}; the disruption generated by Saturn's 2010-2011 northern storm \citep{11fletcher_storm, 11sanchez, 12fletcher, 15moses, 15cavalie}; and the $\sim 15$-year-period oscillatory pattern in the equatorial stratosphere \citep{08orton_qxo, 08fouchet}.  Seasonal gradients therefore cannot be understood in isolation from the atmospheric dynamics, circulation, and chemistry, and we encourage the reader to review this chapter alongside those on atmospheric dynamics (Chapter 12, Flasar \textit{et al.}), polar phenomena (Chapter 13, Sayanagi \textit{et al.}), and upper atmospheric observations (Chapter 9 by Moore \textit{et al.} and Chapter 10 by Koskinen \textit{et al.}).
\index{seasonal asymmetry}

\begin{figure*}
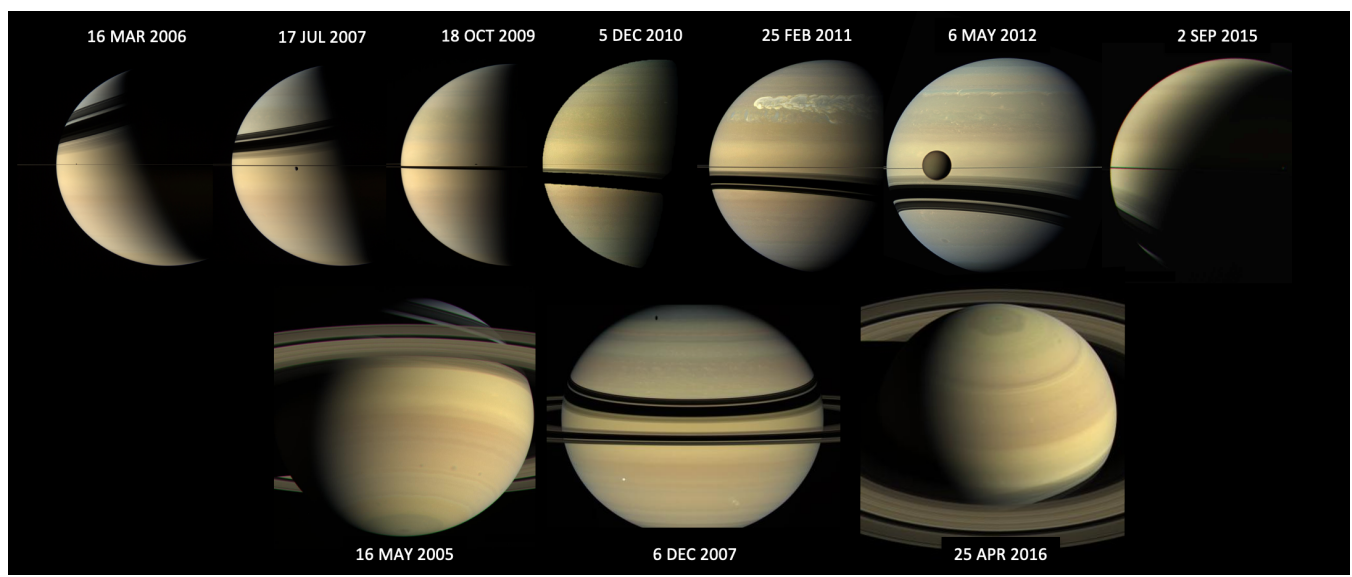
%
\begin{center}
\figurebox{7in}{}{CassiniISSmontage.png}
\caption{Saturn's changing insolation from 2005 to 2016 as seen by Cassini's Imaging Science Subsystem (ISS). The top row shows Saturn from an equatorial vantage point (the rings appear edge-on), and attempts to show the location of the terminator in each image to indicate how it changed over time.  The bottom row shows images from outside of the equatorial plane, showing the transition from southern summer to northern summer.  Compiled from Cassini/ISS images courtesy of NASA/JPL-Caltech.}
\label{saturn_montage}
\end{center}
\end{figure*}

By the end of 2014, it was clear that Saturn's atmospheric temperatures above the radiative-convective boundary (400-500 mbar) were tracking seasonal changes to insolation, and that radiative modelling \citep{90conrath, 12friedson, 14guerlet} was largely successful in predicting the amplitude and phase lag of the seasonal response.  There were some surprises - for example, the mid-stratospheric seasonal response appeared to be smaller than expected \citep{15sylvestre}.  Differences between modelled and measured temperatures are likely driven by the use of spatially-uniform hydrocarbons in the models (key stratospheric coolants), by residual-mean circulation resulting in adiabatic heating and cooling, and also by simplistic assumptions about aerosol contributions to radiative balance.  Indeed, aerosol variability may play a key role, and by the end of 2014, Cassini's visible-light cameras had revealed the shifting colours of Saturn's springtime hemisphere (Fig. \ref{saturn_montage}), from blue hues to the more familiar yellow-ochre appearance.  However, only limited quantitative studies had been performed to assess Saturn's aerosols at single epochs with Cassini \citep{13roman, 11fletcher_vims, 13sromovsky, 20sromovsky}, and no synthesis of seasonal variations had been presented.  Furthermore, there was suggestive evidence of interannual variability in the temperature structure, as comparisons of measurements from Voyager 1 (1980, $L_s=8.6^\circ$) and 2 (1981, $L_s=18.2^\circ$) with Cassini at the same point in the seasonal cycle (2009-10) revealed some unexpected differences, particularly at the equator \citep{13li, 14sinclair, 16fletcher}.  As we show below, the extension of Cassini's lifetime to northern summer solstice and beyond proved to be of great benefit in understanding the timescales for these atmospheric changes. 
\index{radiative-convective boundary}
\index{residual-mean circulation}
\index{interannual variability}

Our previous review of Saturn's seasons therefore left a number of open questions, particularly concerning the expected seasonal evolution between mid-spring and summer in the northern hemisphere.  Although Cassini had tracked atmospheric temperatures over a full decade, the expected development of a warm north polar stratospheric vortex had not yet started \citep{15fletcher_poles}.  Furthermore, there was limited evidence for reversing hemispheric asymmetries in tropospheric and stratospheric composition and aerosols, raising the question of whether some hemispheric asymmetries were permanent.  Neutral photochemistry modelling \citep{05moses_sat, 15hue} had predicted the magnitude of hemispheric contrasts for a range of hydrocarbons, but remaining discrepancies between models and data hinted at the influence of stratospheric circulation.  As we show in this chapter, a more complete dataset of seasonally-changing temperatures and hydrocarbons substantially improves our ability to test the predictions of radiative-convective and photochemical models. 
\index{polar stratospheric vortex}

% (between winter and summer solstice) for a range of hydrocarbons, and although Cassini measurements from single moments in time were prone to substantial uncertainties, they had been able to reveal discrepancies between models and observations that were suggested to be due to the influence of stratospheric circulation.  

This chapter reviews Saturn's seasonal development since 2014, and is organised as follows.  The next Section reviews the ongoing evolution of Saturn's temperatures (Section \ref{temp}), in comparison to expectations from radiative-climate and general circulation models.  Section \ref{chem} then explores seasonal changes in tropospheric and stratospheric composition and compares measurements to the state of the art in seasonal chemical modelling.  Section \ref{aerosols} discusses the observed changes in the properties of Saturn's clouds and hazes.  Finally, Section \ref{conclude} describes the next stages in our exploration of Saturn's evolution in the 2020s.  The reader is referred to \citet{18fletcher_book} for detailed descriptions of Cassini's suite of remote sensing instruments; historical surveys of pre-Cassini observations; radiative-convective model development since the earliest equilibrium models in the 1970s; and the development of photochemical models in the troposphere \citep[e.g., phosphine,][]{09visscher} and stratosphere \citep[e.g., methane,][]{05moses_sat, 15hue}.  Furthermore, this chapter will not dwell on Saturn's bulk chemical composition, for which the reader is also referred to \citet{09fouchet}, nor on the influence of exogenic material on the upper atmosphere, which was reviewed in \citet{18fletcher_book}, and is discussed further in Chapters 9 (Moore \textit{et al.}) and 10 (Koskinen \textit{et al.}) based on new observations by the Cassini INMS instrument \citep{18waite}.  Instead, we focus on those chemical species, atmospheric properties, and numerical models that provide insights into seasonal trends on Saturn near northern summer solstice.

%%%%%%%%%%%%%%%%%%%%%%%%%%%%%%%%%%%%%%%%%%%%%
%%%%%%%%%%%%%%%%%%%%%%%%%%%%%%%%%%%%%%%%%%%%%
%%%%%%%%%%%%%%%%%%%%%%%%%%%%%%%%%%%%%%%%%%%%%
\section{Temperature Variations approaching Summer Solstice}
\label{temp}

% Introduction to radiative balance
At altitudes above the radiative-convective boundary \citep[350-500 mbar,][]{07fletcher_temp}, Saturn's upper tropospheric and stratospheric temperatures are governed by the balance between radiative heating and cooling, to first order.  Heating occurs via short-wave absorption of sunlight by methane and aerosols, whereas stratospheric cooling is due to long-wave emission from ethane, acetylene, and (to a lesser extent) methane and other hydrocarbons in the mid-stratosphere ($p<5$ mbar), and tropospheric cooling is dominated by emission from the hydrogen-helium continuum ($p>5$ mbar).  The amplitude of hemispheric asymmetries depends on the radiative time constant at each altitude \citep[e.g.,][]{79cess, 90conrath} -- when this exceeds a Saturnian year in the troposphere, the thermal field does not show seasonal asymmetries and should be close to the annual-mean radiative equilibrium values \citep[at depths below 500-700 mbar,][]{83conrath}.  When the radiative time constant is very short in the high stratosphere, the temperatures track the sub-solar point and extremes of temperature should be larger (i.e., a low thermal inertia due to having less mass at low pressures).  At intermediate altitudes sensed by Cassini/CIRS, the hemispheric asymmetries in temperature lag behind the instantaneous insolation, sometimes by up to a season \citep[e.g,][]{90conrath, 98conrath}.  However, the picture is made more complicated because the key stratospheric coolants (ethane and acetylene) respond to seasonal photochemistry and are advected by circulation; the thickness, altitude, and reflectivity of tropospheric and stratospheric aerosols vary with season; and atmospheric circulation responds to large-scale temperature gradients by residual mean circulation, adiabatic heating/cooling, and wave activity.  To date, no model combines the feedback processes between all of these parameters, but Cassini observations were able to constrain the temporal evolution of the various environmental parameters, albeit not over a full Saturn year.
\index{radiative-convective boundary}
\index{radiative time constant}

\adjustfigure{60pt}

% Review of 2015 Chapter
The first decade of Cassini observations, reviewed in \citet{18fletcher_book}, tracked Saturn's seasonal temperature evolution beyond the first snapshots of hemispheric asymmetries \citep{04flasar, 07fletcher_temp, 08fletcher_poles, 09guerlet}.  These early studies revealed 1-mbar stratospheric temperature contrasts of $\sim40$ K between the summer and winter poles, decreasing to $\sim10$ K at the tropopause. Subsequent work revealed the cooling of the southern autumn hemisphere and the warming of the northern spring hemisphere \citep{10fletcher_seasons, 10li, 13sinclair, 15fletcher_poles}, leading to almost symmetric mid-latitude 1-mbar temperatures in the 140-145-K range by equinox.  Seasonal changes were particularly apparent with the cooling and disappearance of a warm south-polar stratospheric vortex (SPSV) that had been present in the $75-90^\circ$S region at the start of the Cassini mission \citep[sometimes referred to as a `polar hood,'][]{05orton, 08fletcher_poles, 15fletcher_poles}. At higher stratospheric altitudes ($\sim0.01$ mbar), a comparison of Cassini/CIRS limb measurements between \citet{09guerlet} and \citet{15sylvestre} suggested that seasonal trends and asymmetries were less significant than those observed at 1 mbar, counter to the expectations of radiative equilibrium models.   
% Remote sensing observations were supported by radio occultations at some latitudes and times \citep[e.g.,][]{11schinder}, but CIRS data have dominated the seasonal studies, as discussed below.
\index{temperature}
\index{polar stratospheric vortex}
\index{CIRS}
\index{seasonal asymmetry}

\begin{figure}%
\begin{center}
\figurebox{3.5in}{}{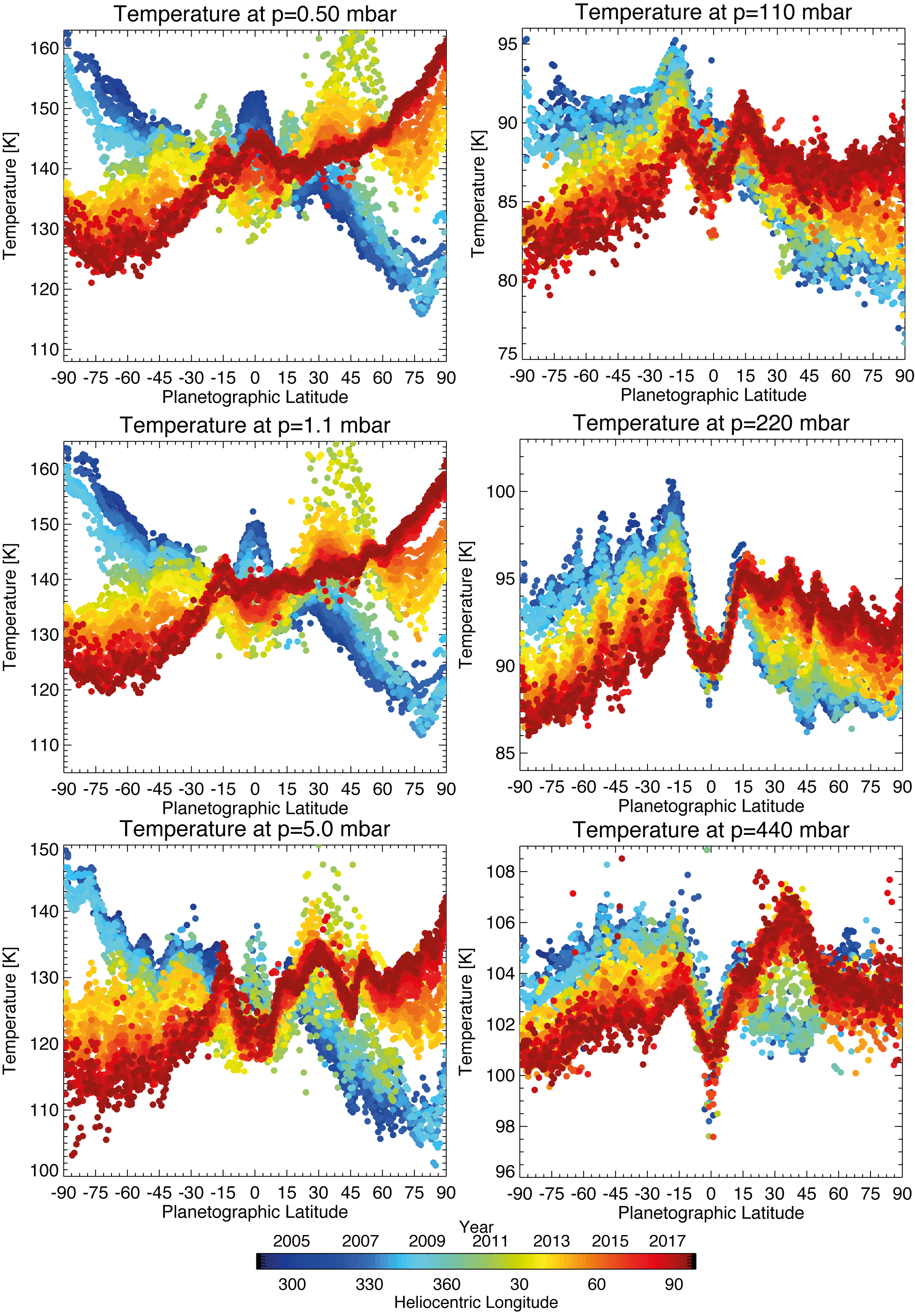}
\caption{Stratospheric (left) and tropospheric (right) temperatures derived from fitting Cassini/CIRS nadir observations using both the far-infrared (16-1000 $\mu$m) and mid-infrared (7-16 $\mu$m) focal planes \citep{16fletcher, 18fletcher_poles}.  Spectra were averaged on a monthly grid and are colour-coded according to the legend, showing the evolution from the start of Cassini's mission (2004) to the end (2017).  Note the influence of Saturn's equatorial oscillation in the stratosphere, and the perturbations due to the 2010-2013 storm-induced stratospheric `beacon.'  }
\label{nadirTresults}
\end{center}
\end{figure}

% Discussion of temperature changes from CIRS
As Saturn approached northern summer solstice in May 2017 ($L_s=90^\circ$), Cassini/CIRS continued to monitor the thermal field.  A series of papers utilised CIRS nadir spectra at the lowest spectral resolution (15 cm$^{-1}$), using tensioned splines to interpolate and `reconstruct' Saturn's temperatures throughout the full 13-year mission.  This technique was used to investigate tropospheric seasonal change using far-IR nadir spectra \citep[16 $\mu$m to 1 mm,][]{16fletcher}; equatorial stratospheric oscillations \citep{17fletcher_QPO} and polar vortices \citep{18fletcher_poles} using mid-IR nadir spectra (7-16 $\mu$m).  The full set of CIRS nadir temperatures in the troposphere (100-440 mbar) and stratosphere (0.5-5.0 mbar) are presented in Fig. \ref{nadirTresults}, where each point represents an inversion of a 7-200 $\mu$m spectrum averaged for a particular latitude and month.  By northern summer solstice in 2017, the stratospheric temperatures (0.5-1.0 mbar) at high latitudes had warmed by 35 K, such that the asymmetries observed at the start of Cassini's mission in 2004 had almost completely reversed.  However, at 5 mbar in Fig. \ref{nadirTresults} the northern-solstice asymmetry ($\sim20$ K) remained smaller than that seen in 2004 ($\sim35$ K).  The north polar region continued to warm beyond northern solstice at $L_s=90^\circ$ \citep{22blake}, so it is likely that the asymmetry had not reached its maximum in 2017.  Furthermore, the different asymmetry amplitude between northern and southern solstice could be related to the small differences between perihelion (near $L_s=280^\circ$) and aphelion (near $L_s=100^\circ$) \citep[small 2-3 K differences are observed at 0.5 mbar in radiative climate models, see below and][]{16hue}.  This conundrum was also observed in the troposphere (110-220 mbar), where gradients reversed \citep[and were largely symmetric by 2013-15,][]{16fletcher}, but the northern hemisphere would continue to warm beyond the end of Cassini's mission.  Finally, at 440 mbar (close to the radiative-convective boundary) we see seasonal changes of 4-5 K or smaller, with much of the observed variability in Fig. \ref{nadirTresults} driven by dynamics (see below).
\index{stratospheric oscillation}
\index{temperature}
\index{radiative-climate model}

% Looking at Tommy's model output (without the photochemical model feedback), the north/south temperature asymmetry (at +/-89˚ latitude) at 0.5mbar and Ls=270˚ is 42K, while it is 39.4K at Ls=90˚

\begin{figure*}%
\begin{center}
\figurebox{7.0in}{}{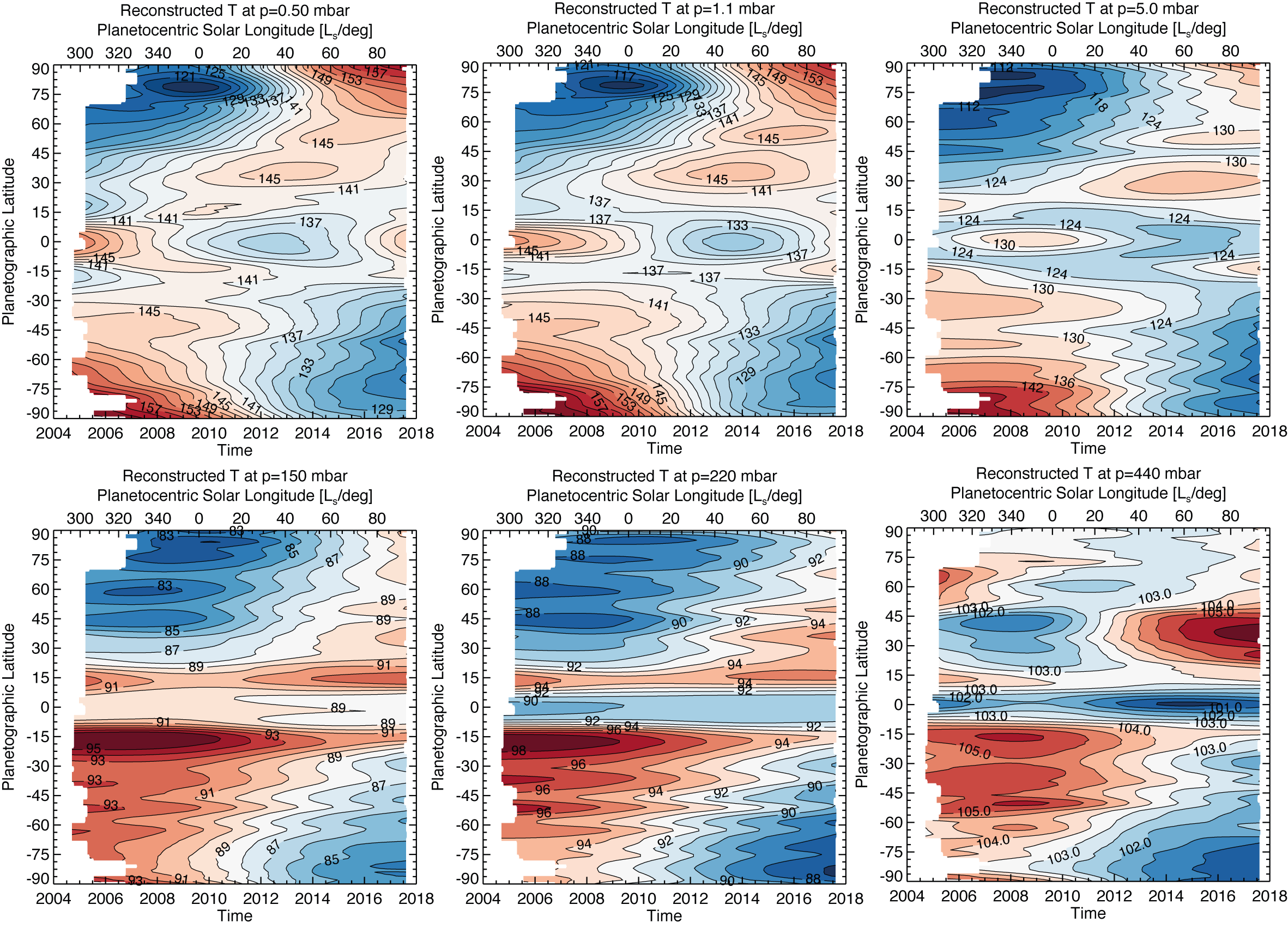}
\caption{Saturn's stratospheric and tropospheric temperatures reconstructed from the measurements in Fig. \ref{nadirTresults} using tensioned splines,  extending the figures in \citet{18fletcher_poles} to cover all latitudes.  Whitespace at the start and end of the time series show where the measurements are unavailable (i.e., splines were used for interpolation but not extrapolation).  We omitted the influence of the 2011-2013 storm at mid-latitudes in the north.}
\label{reconstr_temp}
\end{center}
\end{figure*}

% Discussion of polar changes
The reconstructed temperature fields are shown in Fig. \ref{reconstr_temp}, revealing seasonal behaviour at both poles, switching from a cold polar vortex in winter to a warm polar vortex in summer.  These figures omit the strong rise in temperatures associated with Saturn's great northern storm that can be seen in Fig. \ref{nadirTresults} \citep{12fletcher, 14achterberg}, whose aftermath could still be observed in the warm troposphere at 440 mbar even in 2017 (but not in the radiatively-controlled upper troposphere at $p<400$ mbar).  The equatorial stratospheric oscillation is superimposed onto the seasonal temperature changes evident in Fig. \ref{reconstr_temp}.  The banded thermal structure is apparent in the troposphere and stratosphere at 5 mbar, but less so at the 0.5-1.0 mbar level.

The changing latitudinal temperature gradients in Fig. \ref{reconstr_temp} are related to the vertical shear of the zonal winds via the thermal wind equation \citep[e.g.,][]{04holton}.  In the stratosphere, the seasonal change in the temperature gradient (from decreasing with northward latitude to increasing with northward latitude) promotes increasingly positive vertical shear on zonal jets in the southern autumn/winter stratosphere, and increasingly negative vertical shear on zonal jets in the northern spring/summer stratosphere \citep{12friedson, 18fletcher_poles}.  If we assume a level of zero winds somewhere in the lower stratosphere, the strengthening negative windshear in the north would promote westward winds surrounding the forming north polar stratospheric vortex (NPSV), particularly at 5 mbar near $78^\circ$N after 2014.  The NPSV also exhibited a hexagonal boundary in the stratosphere, mirroring that in the troposphere \citep{18fletcher_poles}.  Similarly, the weakening of the westward winds in the south (particularly near $74^\circ$S) corresponded to the disappearance of the SPSV after 2013-14.  Ground-based imaging revealed that the NPSV continued to warm (therefore strengthening the circumpolar westward windshear) between 2017 and 2022 \citep{22blake}.   
\index{Great White Storm}
\index{polar stratospheric vortex}
\index{thermal wind equation}
\index{zonal wind}

%\citet{16fletcher} described the implications of the changing tropospheric temperature gradients on :  seasonal changes were smaller than 10 m/s per scale height near the tropopause, larger in the northern hemisphere than in the south, and primarily affected the broad retrograde jets, rather than the narrower prograde jets.    

\subsection{Simulating Seasonal Variability}

\begin{figure*}%
\begin{center}
\figurebox{7.0in}{}{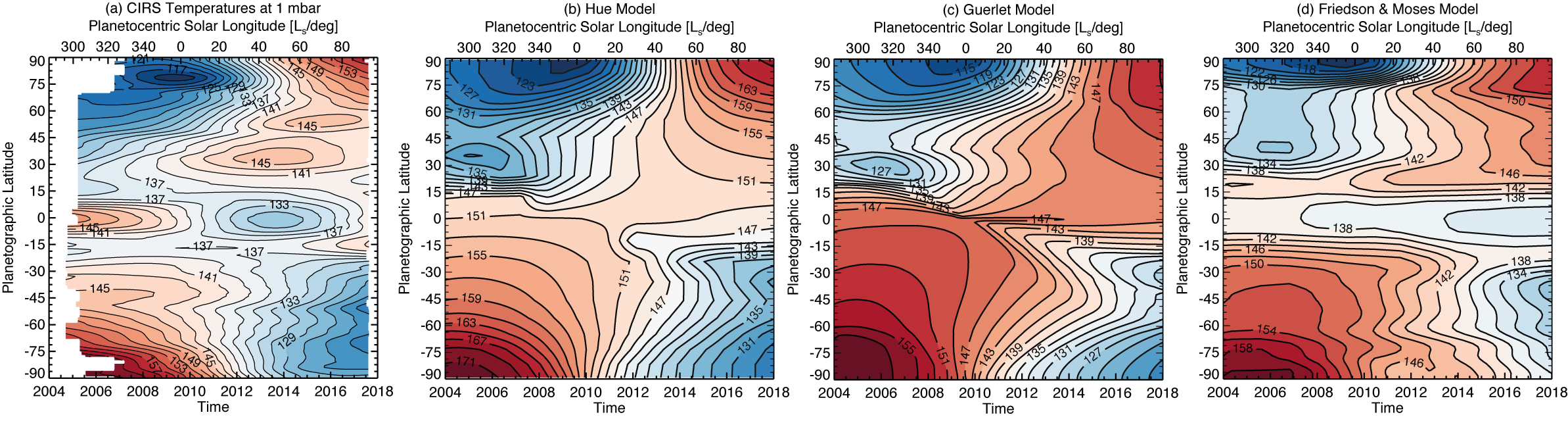}
\caption{Comparison of the reconstructed 1-mbar temperature field from Fig. \ref{reconstr_temp} with simulations from \citet{16hue}, \citet{14guerlet} and \citet{12friedson}.  Only the latter model includes the influence of atmospheric circulation, but all three are successful in reproducing the seasonal asymmetries and the approximate timings of maxima and minima in temperature.  }
\label{modelcomp}
\end{center}
\end{figure*}

\begin{figure}%
\begin{center}
%\vspace{80pt}
\figurebox{3.0in}{}{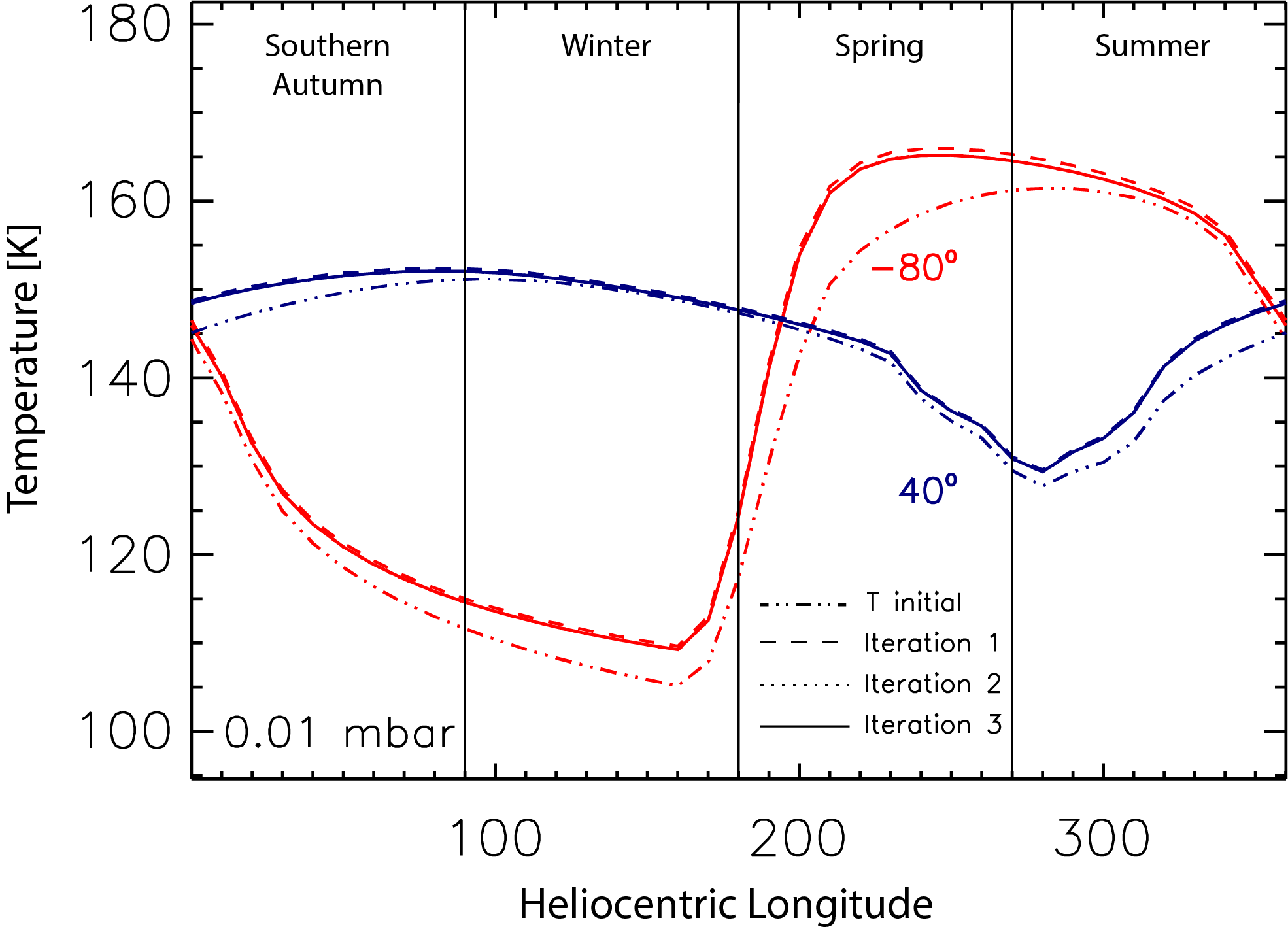}
\caption{Seasonal evolution of temperature at 0.01 mbar and latitudes of 40$^{\circ}$N (blue) and 80$^{\circ}$S (red). The multiple iterations between the photochemical and radiative seasonal models of \citet{16hue} are shown as different lines (dash double dotted line: initial state, dashed lines: first iteration, solid line: final iteration).}
\label{fig:hue2016_Fig2}
\end{center}
\end{figure}

\textit{How well are these seasonal changes captured by radiative-convective (RC) and global climate models?} Fig. \ref{modelcomp} compares the CIRS 1-mbar temperature field with predictions from three numerical models of differing complexity.  \citet{12friedson} described the Outer Planet General Circulation Model (OPGCM), which tracks radiative heating and cooling and accounts for dynamical redistribution of heat via adiabatic expansion/compression. \citet{14guerlet} developed the RC model that was incorporated into the GCM of \citet{20spiga}, itself extended to the upper stratosphere in \citet{21bardet} --- we show the RC estimates of seasonal temperatures here.  In both cases, the spatial distributions of hydrocarbons were treated as latitudinally and temporally uniform, and the influence of stratospheric aerosols, which can cause increased stratospheric temperatures of 5-6 K \citep{15guerlet}, was not considered.  Going one step further, \citet{16hue} coupled the radiative model of \citet{10greathouse} with the photochemical model of \citet{15hue}, showing how seasonal variations of the hydrocarbons would influence the thermal field (and vice versa), but without coupling to a full dynamical model.  The main effect was on the thermal field at high latitudes (Figs. \ref{modelcomp} and \ref{fig:hue2016_Fig2}), as seasonal chemistry and transport affects the distribution of stratospheric coolants. 
\index{radiative-climate model}
\index{general circulation model}
\index{photochemistry}

\adjustfigure{60pt}

Despite these differing complexities, all three models effectively reproduce the CIRS trends, once the equatorial stratospheric oscillation is discounted, confirming the dominance of radiative heating and cooling in shaping Saturn's temperatures.  At 1 mbar in Fig. \ref{modelcomp}, the 6-to-8-year phase lag between minimum insolation (winter) and the coolest stratospheric temperatures is correctly predicted by all of the models, reproducing the timings of temperature maxima and minima to the extent allowed by the 13-year span of the CIRS data. The results of \citet{16hue} suggest that this replication of the phase lags can be improved still further when the meridional and temporal variation in stratospheric coolants is also included, ideally via coupling to a circulation model, rather than holding them fixed with latitude and time.  At higher altitudes (0.01 mbar) in the spring hemisphere (Fig. \ref{fig:hue2016_Fig2}), radiative modelling suggests that as the insolation level increases, the stratosphere heats up with limited cooling due to the low abundances of photochemical by-products of methane at that time. Later during that season, as the amounts of C$_2$H$_2$ and C$_2$H$_6$ increase (see Section \ref{chem}), the atmospheric cooling rates balance out the heating and the temperature decreases. This produces a 0.01-mbar temperature peak before the $L_s=270^\circ$ solstice in Fig. \ref{modelcomp} at $80^\circ$S in Fig. \ref{fig:hue2016_Fig2} \citep{16hue}.  
% This early peak is increasingly more pronounced at high altitude and latitude -- this effect is illustrated in Fig. \ref{fig:hue2016_Fig2}, which shows the seasonal evolution of the temperature at 0.01 mbar.  
\index{phase lag}
\index{radiative-convective model}

Quantitative differences in absolute temperatures between the models are likely related to the differing hydrocarbon assumptions \citep[detailed in Section 10.2.4 of][]{18fletcher_book}, and potentially to the absence of latitudinally- and temporally-variable aerosols.  \citet{14guerlet} calculated that the radiative contribution from minor species such as C$_3$H$_8$, CH$_3$C$_2$H, C$_4$H$_2$ and CH$_3$D would be about 5\% of the total cooling rate, and therefore neglected them in their model.  Inclusion of both dynamical redistribution of energy and hydrocarbons \citep[particularly in the location of ring shadowing,][]{09guerlet}, alongside seasonal photochemistry and aerosols, may eventually close the gap between the models and the observations.  Indeed, where cool regions under the ring's shadow were observed in the RC model of \citet{14guerlet}, dynamical heating was produced in the GCM of \citet{21bardet}.  A similar result was obtained in the circulation model of \citet{12friedson}, with subsidence (of order 0.2 mm/s at 1 mbar) and warming at low latitudes in winter, removing this cold thermal anomaly.  At high latitudes, \citet{15fletcher_poles} suggested that vertical velocities (and thus adiabatic warming/cooling) as small as 0.1 mm/s could account for discrepancies between observations and these simple equilibrium models at the 1-mbar level.  
\index{ring shadow}
\index{general circulation model}
\index{residual-mean circulation}
\index{radiative-climate model}
\index{circulation}

\adjustfigure{60pt}

\begin{figure}%
\begin{center}
\figurebox{3.0in}{}{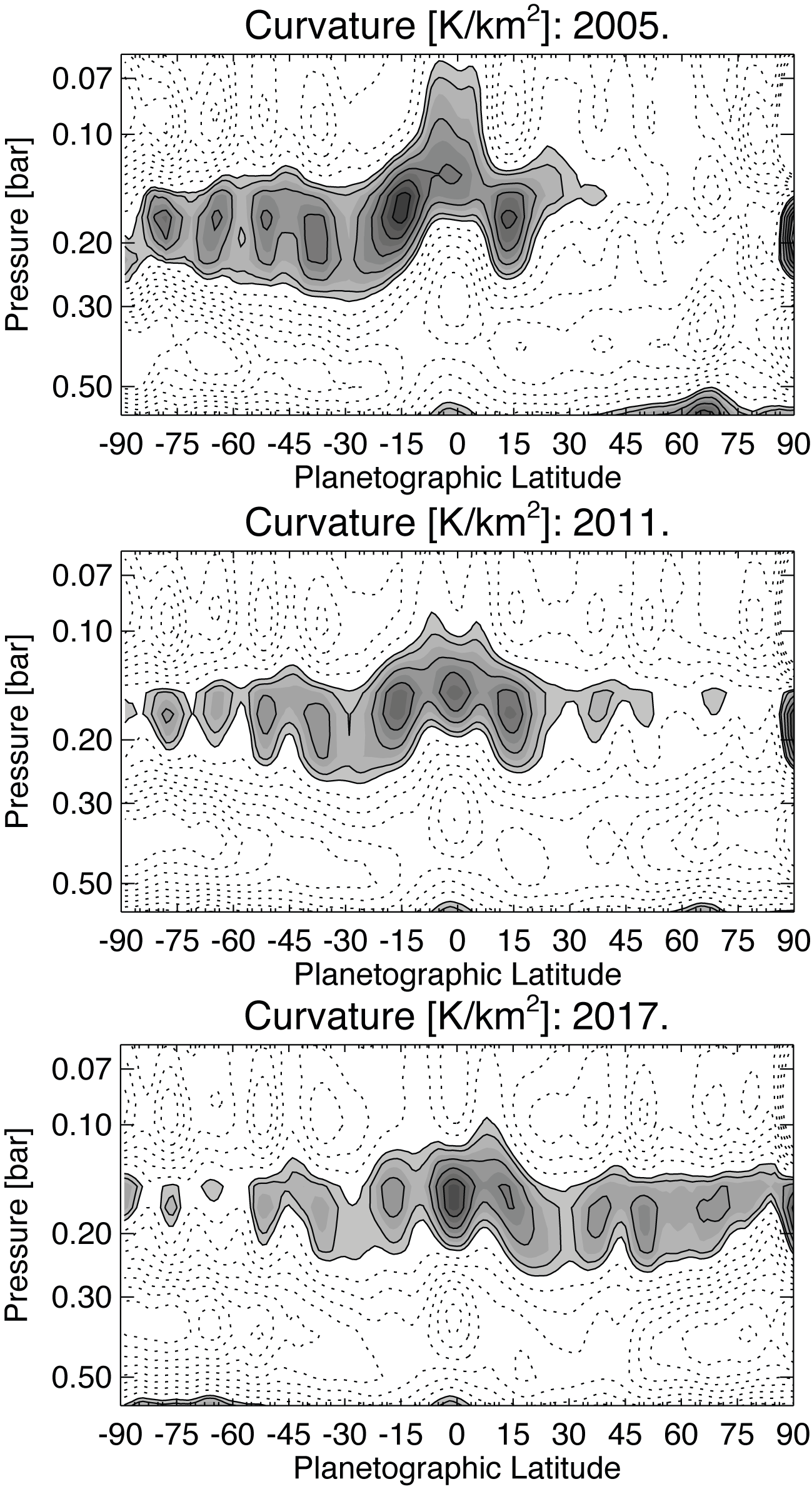}
\caption{Seasonal progression of the temperature knee from \citet{16fletcher}, extended to 2017.  The term `curvature' refers to the vertical gradient of the lapse rate.  Contours are spaced every 0.004 K/km$^2$, with grey shading and solid lines used to indicate the strongest negative curvature (i.e., the strongest temperature knee), and dotted lines showing positive curvature.  The influence of the stratospheric oscillation can also be seen in the equatorial region.}
\label{knee}
\end{center}
\end{figure}

% Temperature Knee
Aerosols have a direct and measurable impact on Saturn's tropospheric temperature, producing localised heating between the radiative-convective boundary and the tropopause (sometimes referred to as the `temperature knee') that was enhanced in southern summer, weak or absent in northern winter, and elevated over the equator \citep{07fletcher_temp, 81hanel, 85lindal}.  This localised heating was also observed in the models of \citet{12friedson} and \citet{14guerlet}, confirming it to be a radiative effect, and \citet{14guerlet} suggested that larger tropospheric aerosols would increase the size of the thermal perturbation.  \citet{16fletcher} tracked the curvature of Saturn's tropospheric temperature profiles (i.e., the derivative of the lapse rate), and showed the perturbation grew smaller in the southern autumn and began to extend northwards in the spring hemisphere by 2014.  Fig. \ref{knee} extends this analysis to the end of the Cassini mission, showing the continued growth in the springtime hemisphere due to increased insolation \citep{12friedson} and, possibly, temporal changes in the aerosol content \citep{05karkoschka}.  
\index{temperature knee}
\index{aerosols}

\subsection{Connecting Stratosphere and Thermosphere}

\textit{How do the mid-stratospheric temperatures connect to the thermosphere?} The shifting hemispheric contrasts and polar vortices are primarily driven by radiative balance and atmospheric circulation in the stratosphere, but here we note potential connections to higher altitudes.  At nanobar pressures, far above the stratospheric domain sensed by CIRS, exospheric temperatures increase to 400-500 K \citep{15koskinen}.  Saturn's main auroral ovals \citep{11badman} appear to encompass the stratospheric polar vortices \citep{18fletcher_poles}, suggesting possible links between particle precipitation, energy deposition, and potentially ion-neutral chemistry at high latitudes within the polar vortices.  The thermosphere is deeper at the poles, leading to substantially higher temperatures than at lower latitudes in the lower thermosphere, reminiscent of the warm vortices observed in the stratosphere.   However, a pole-to-pole map of upper thermospheric temperatures from UVIS stellar occultations \citep{20brown} suggests that temperatures peak near 60-$70^\circ$ latitude and then decline towards each pole.  Furthermore, energy is redistributed through the thermosphere equatorward from the auroral regions, potentially due to gravity wave drag enabling meridional circulation \citep{22brown}.  The cool poles are counter to the Saturn Thermosphere Ionosphere General Circulation Model \citep[STIM,][]{19muller-wodarg} simulations, which suggests that the warm regions poleward of $80^\circ$ latitude for the lower thermosphere should also persist at nanobar pressures.  This discrepancy remains a topic of active research.  Furthermore, the STIM model predicts warmer summertime thermospheric temperatures due to enhanced resistive (Joule) heating -- this is due to larger solar ionisation and ionospheric plasma densities, rather than due to direct solar heating.  However, to date the UVIS occultation data do not show evidence for seasonal asymmetries in temperature \citep{20brown}.  This could be a consequence of the effective northward offset of Saturn's magnetic field \citep{18dougherty}, with the weaker southern field producing a wider area of high Joule heating rates in the south \citep{06muller-wodarg} that balance the enhanced Joule heating in the northern summer hemisphere.  Connecting these processes at nanobar pressures to millibar and microbar pressures remains a significant modelling challenge, and future progress may be made via comparison to Jupiter's time-variable auroral heating \citep{21odonoghue}, chemistry \citep{19sinclair}, aerosol production \citep{13zhang_aer}, and high-altitude winds \citep{18johnson, 21cavalie}.

\index{thermosphere}
\index{Joule heating}
\index{UVIS}
\index{occultation}

% The issue of the magnetic field is discussed by, for example, Muller-Wodarg et al. (2006), Section 4.3: "The asymmetries between north and south are due to the quadrupole terms in the Saturn Pioneer Voyager (SPV) model of internal magnetic field (Davis and Smith 1990), which results in north-south differences of the magnetic field strength, affecting the Joule heating rates. While northern peak rate is larger, the region is wider in the south..." Effectively, the weaker magnetic field in the south leads to a larger area of Joule heating in the south. A back of the envelope argument is presented, for example, by O'Donoghue et al. (2014, Icarus, 229, 214), section 4.1. The new magnetic field from Cassini is in Dougherty et al. (2018), for our purposes identical to SPV. Dougherty et al. (2018) write: "Saturn's magnetic equator, where the magnetic field becomes parallel to the spin axis, is shifted northward by 2808.5 km." 

Saturn's temperatures have continued to evolve beyond northern summer solstice, and it will be the role of ground-based \citep{17guerlet_dps, 22blake} and space-based \citep{16norwood} infrared observations to extend the time series, as discussed briefly in Section \ref{conclude}. 

%%%%%%%%%%%%%%%%%%%%%%%%%%%%%%%%%%%%%%%%%%%%%%%%%%%%%%%%%%%%%%%
%%%%%%%%%%%%%%%%%%%%%%%%%%%%%%%%%%%%%%%%%%%%%%%%%%%%%%%%%%%%%%%
%%%%%%%%%%%%%%%%%%%%%%%%%%%%%%%%%%%%%%%%%%%%%%%%%%%%%%%%%%%%%%%
\section{Seasonal Variations in Atmospheric Composition}
\label{chem}

Whereas the seasonal changes in tropospheric and stratospheric temperatures have been straightforward to deduce from Cassini remote sensing, the changes to gaseous abundances are extremely challenging to measure due to degeneracies inherent in spectral inversions.  In this section we review changes to spatial distributions of chemicals as Saturn approached northern summer solstice, starting in the troposphere and moving into the stratosphere.
\index{composition}

\subsection{Tropospheric Composition Variations}

% Phosphine and Ammonia
An overview of Saturn's tropospheric composition, including both condensation processes and photochemistry, can be found in \citet{09fouchet} and \citet{18fletcher_book}.  Of the key volatile species (NH$_3$, H$_2$S, H$_2$O), only ammonia has been mapped as a function of latitude, revealing contrasts between the equatorial zone and neighbouring belts indicative of equatorial upwelling \citep{11fletcher_vims, 13laraia, 13janssen, 16barstow}, and hints of a subtle north-south asymmetry \citep[with less NH$_3$ in the cooler northern hemisphere during winter, potentially due to condensation,][]{11fletcher_vims, 12hurley}.  Of the disequilibrium `quenched' species (CO, PH$_3$, GeH$_4$, AsH$_3$) that trace mixing from the deeper troposphere, only arsine and phosphine have been mapped with latitude, and both show moderately elevated abundances in the southern summer compared to northern winter \citep{09fletcher_ph3, 11fletcher_vims}.  As the vertical profile of PH$_3$ is sensitive to UV photolysis \citep{84kaye, 09visscher}, this may be evidence of UV shielding by increased aerosol opacity in the southern summer hemisphere \citep{18fletcher_book}, although we caution that abundances extracted from VIMS 5-$\mu$m observations are somewhat degenerate with the assumed aerosol properties.  To date, no seasonal studies of NH$_3$, PH$_3$, or AsH$_3$ have been performed, and models that predict the seasonal variation of tropospheric photochemistry on Saturn have not yet been developed.  The question of whether the subtle asymmetries are permanent or time-variable remains unanswered, and the feedback effects of aerosol- and self-shielding (i.e., reducing the rates of photolytic destruction) remain unexplored.  We discuss changes in the high-latitude distribution of PH$_3$, and the putative diphosphine (P$_2$H$_4$) layer, in Section \ref{aerosols}.
\index{tropospheric composition}
\index{phosphine}
\index{ammonia}
\index{VIMS}
\index{photochemistry}
\index{arsine}
\index{diphosphine}
\index{disequilibrium species}

% Julie:  It might be useful comparing the PH3 and AsH3 distributions with what was determined for Jupiter from JIRAM (Grassi et al. 2020)? 

\begin{figure*}%
\begin{center}
\figurebox{6.0in}{}{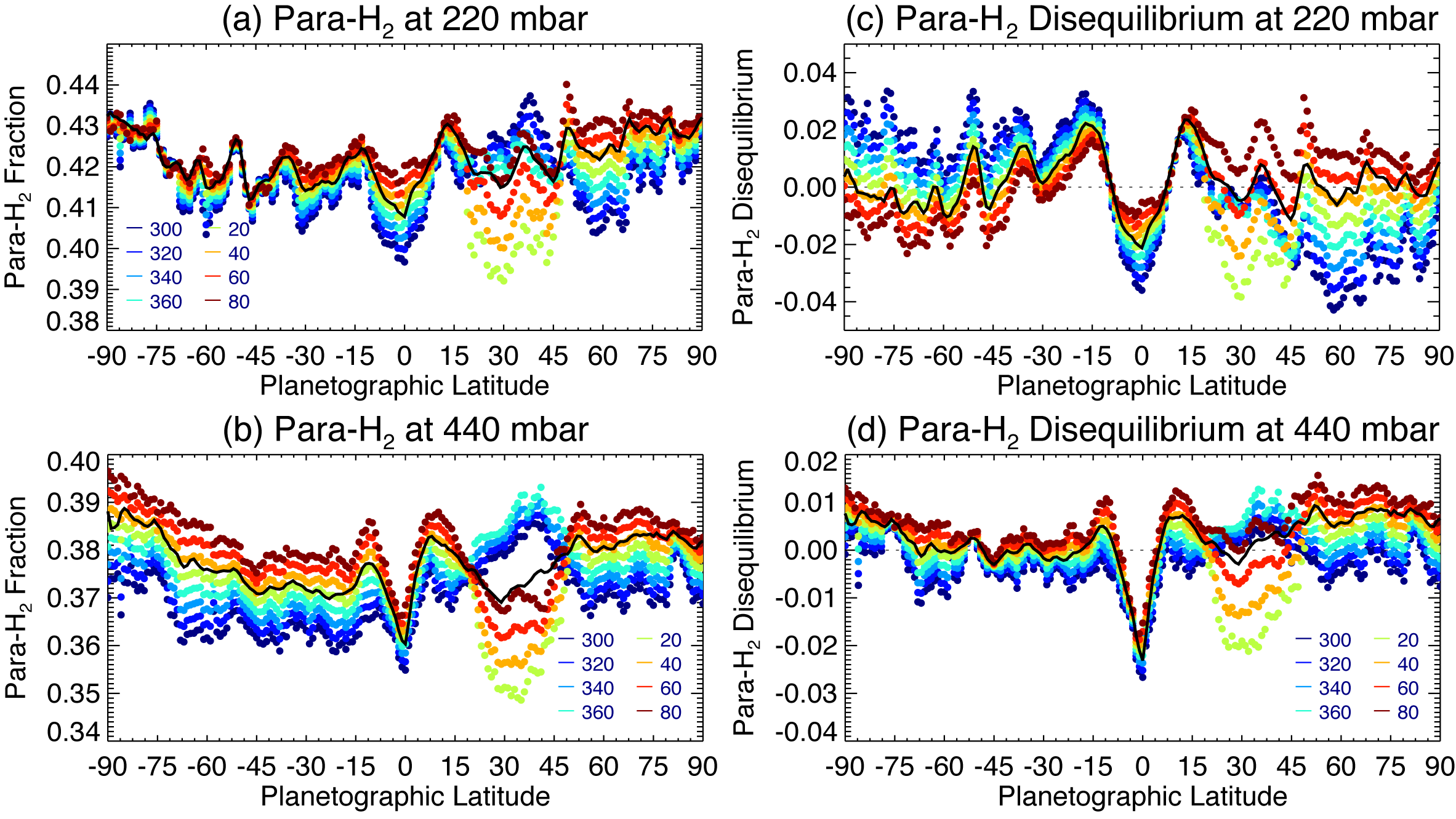}
\caption{Latitudinal distribution of para-H$_2$ fraction (left) and extent of disequilibrium (right) at 220 and 440 mbar, updated from \citet{16fletcher}.  Distributions at different values of $L_s$ are extracted from a linear interpolation to the full CIRS time series, with a two-step linear fit used for latitudes perturbed by the 2010-11 northern storm \citep{14achterberg}.  Solid black lines represent the mean of all inversions during the 13-year mission.  Dotted horizontal black lines in the disequilibrium plots denote equilibrium conditions.  Typical uncertainties on $f_p$ inversions are $\sim\pm0.01$ \citep{16fletcher}.}
\label{parah2}
\end{center}
\end{figure*}

More progress has been made in the study of Saturn's para-hydrogen distribution, which traces both mixing from the deeper troposphere and the efficiency of interconversion between spin isomers of H$_2$ \citep[catalysed by the presence of paramagnetic sites on aerosols with equilibration timescales of 90-160 years at 200 mbar,][]{03fouchet}.  The shape of the far-infrared collision-induced continuum is determined by the ratio of ortho-H$_2$ (odd rotational spin isomer with parallel spins governing the broad S(1) feature near 587 cm$^{-1}$) and para-H$_2$ (even spin isomer with anti-parallel spins governing the broad S(0) feature near 354 cm$^{-1}$) \citep{82massie}.  At high temperatures in the deeper troposphere, the fraction of para-H$_2$ ($f_p$) makes up 25\% of the molecular hydrogen, whereas the equilibrium fraction increases to 45-50\% at the cold tropopause.  Displacement of air parcels from the deeper troposphere at a rate faster than the ortho-to-para conversion means that the low para-H$_2$ fraction is `frozen in'.  These `sub-equilibrium' conditions indicate uplift from deeper, warmer levels, which seems to be the case at the equator \citep{07fletcher_temp, 16fletcher} and at the latitudes of Saturn's 2010-11 northern storm \citep{14achterberg}.  Conversely, `super-equilibrium' can indicate sinking from higher, cooler altitudes near the tropopause.   The extent of this disequilibrium is seasonal:  \citet{98conrath} reported super-equilibrium conditions and higher $f_p$ in the springtime hemisphere ($30^\circ$S to $70^\circ$N) during the Voyager era, \citet{07fletcher_temp} showed that this had reversed by southern summer observed by Cassini, with sub-equilibrium in the winter hemisphere polewards of $30^\circ$N.  \citet{16fletcher} showed that super-equilibrium conditions moved northwards as winter turned to spring (see Fig. \ref{parah2}c, right), and sub-equilibrium conditions emerged in southern autumn.  Although small changes in $f_p$ were occurring, much of this asymmetry in disequilibrium can be simply attributed to changing temperatures (and thus the change in the equilibrium para-H$_2$ fraction), without invoking large-scale circulation or changes in aerosol catalysis.  
% However, small changes to $f_p$ were also occurring -- $f_p$ was found to have increased in the north between 2004 and 2014, possibly as a consequence of subsidence in the springtime hemisphere, whereas $f_p$ in the south was largely unaltered.
\index{para-hydrogen}
\index{super-equilibrium}\index{sub-equilibrium}\index{CIRS}\index{circulation}\index{disequilibrium species}

In Fig. \ref{parah2} we extend the $f_p$ mapping from far-IR inversions \citep{16fletcher} to 2017.  This confirms that $f_p$ was relatively stable with time at 220 mbar, and shows the increase in $f_p$ in the northern hemisphere throughout the mission \citep{16fletcher}, only perturbed in the $15-45^\circ$N range by the low-$f_p$ air brought upwards by the 2010-11 storm.  At 440 mbar $f_p$ showed an asymmetry pre-equinox with lower $f_p$ in the summer hemisphere than in the winter.  A similar asymmetry was observed by Voyager in early northern spring \citep{98conrath, 16fletcher}.  CIRS results in Fig. \ref{parah2}b reveal that this asymmetry begins to reverse and become more symmetric after equinox as $f_p$ increased faster in the mid-to-high southern latitudes than anywhere else, although this interpretation is significantly complicated by the influence of the 2010-11 storm.  This can be compared to the expectations for an inter-hemispheric meridional circulation at Saturn \citep{90conrath} --- a two-cell circulation pattern near equinox would lead to subsidence at high latitudes in both hemispheres (their Fig. 10), but approaching solstice the subsidence in the southern winter hemisphere would be stronger (their Fig. 9), rapidly increasing $f_p$ in the south.  With the demise of Cassini, continued monitoring of Saturn's para-H$_2$ will be challenging --- will it continue to exhibit latitudinal asymmetries as indirect evidence of seasonally changing interhemispheric circulation?  The spatial distribution of $f_p$ in the stratosphere remains completely unknown. Furthermore, there is currently no predictive seasonal model that combines the effects of meridional circulation, local dynamical mixing (e.g., at the equator), and the catalytic effects of aerosols on the conversion between ortho- and para-H$_2$.
\index{seasonal asymmetry}\index{para-hydrogen}\index{circulation}

% Models that predict the seasonal variation of tropospheric photochemistry on Saturn have not yet been developed.  However, \citet{16wang} recently studied the chemical kinetics and deep quenching of PH$_3$, AsH$_3$, SiH$_4$, and GeH$_4$ in Saturn’s troposphere.  Latitude variations in the upper-tropospheric mixing ratios of some of these quenched disequilibrium species (especially GeH$_4$), along with quenched CO and C$_2$H$_6$ from the deep troposphere, are expected to result from latitude variations in the eddy diffusion coefficient \citep[e.g.,][]{15wang}.  \citet{16wang} discuss the implications of their modelling for a potential future entry-probe mission to Saturn, including whether quenched disequilibrium species can help constrain the deep water abundance in the likely event that entry probes cannot survive deep enough to measure it directly.
% \index{photochemistry}\index{phosphine}\index{disequilibrium species}\index{tropospheric composition}

\subsection{Stratospheric Composition Variations}

The photolysis of methane in Saturn's upper stratosphere gives rise to a complex network of hydrocarbon species, including C$_2$ species (ethane, acetylene, ethylene), C$_3$ species (propane, methylacetylene) and higher-order hydrocarbons (diacetylene, benzene, etc.). The photochemical reactions responsible for producing these species were reviewed previously \citep{09fouchet, 18fletcher_book}, alongside the published measurements of hydrocarbon abundances from Voyager, Cassini, and ground-based observations before 2014.  Cassini's contribution was to measure the latitudinal distributions of stratospheric species at various points during the 2004-2017 time series.  CIRS observations of ethane (C$_2$H$_6$) and acetylene (C$_2$H$_2$) have been sufficient to trace seasonal evolution \citep{05howett, 09hesman, 09guerlet, 10guerlet, 13sinclair, 14sinclair, 15sylvestre, 15fletcher_poles, 18fletcher_poles}, but before we discuss those findings, we briefly mention the observed distributions of the higher-order hydrocarbon species.  
\index{stratospheric composition}
\index{methane}\index{photochemistry}
\index{ethane}\index{acetylene}\index{CIRS}

Propane (C$_3$H$_8$) at 1 mbar appears to track the instantaneous solar insolation \citep{09guerlet}, but no changes were observed between 2005-06 and 2010 \citep{15sylvestre}.  Ethylene (C$_2$H$_4$) has only been observed in the warm-temperature aftermath of the 2010-11 storm \citep{12hesman, 15moses}, and its latitudinal distribution remains unknown. Methylacetylene (C$_3$H$_4$) and diacetylene (C$_4$H$_2$) were depleted in southern summer mid-latitudes compared to northern winter mid-latitudes, potentially indicating stratospheric upwelling of hydrocarbon-poor air during the southern summer and subsidence of hydrocarbon-rich air in northern winter \citep{10guerlet}.  Benzene (C$_6$H$_6$), measured by both CIRS limb measurements \citep{15guerlet} and UVIS stellar occultations \citep{16koskinen}, appears to be at least as abundant at the southern summer pole compared to the equator, suggesting a role for ion-neutral chemistry in producing benzene at high latitudes (see Section \ref{chemmodel}).  At the time of writing, work continues to study the temporal variability of these species as Saturn approached northern summer solstice.
\index{propane}\index{ethylene}\index{methylacetylene}
\index{benzene}\index{diacetylene}
\index{UVIS}\index{occultation}\index{CIRS}

\begin{figure}%
\begin{center}
\figurebox{3.0in}{}{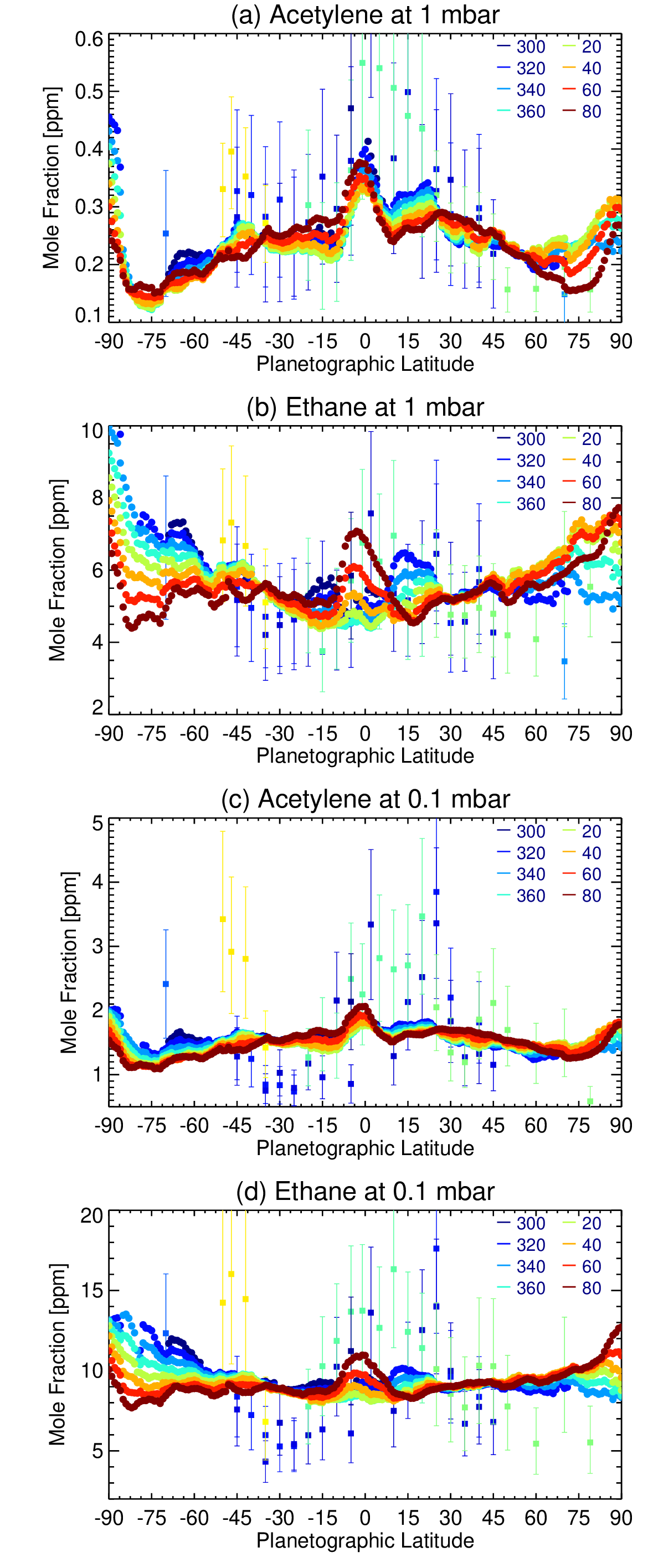}
\caption{Cassini CIRS nadir and limb results for acetylene and ethane at 1 mbar and 0.1 mbar, adapted from \citet{18fletcher_poles} (nadir results as circles, reconstructed from smooth spline fits to inversions of 15-cm$^{-1}$ resolution spectra and extended to all latitudes), \citet{09guerlet} (limb results in 2005-08 as blue-green squares before $L_s=360^\circ$) and \citet{15sylvestre} (limb results in 2010-12 as yellow-orange squares after $L_s=360^\circ$). Note that nadir data do not have sensitivity to the 0.1-mbar level, which is why latitudinal contrasts are much smaller than those observed in the limb data - the observed nadir structure represents the influence of information content in the 0.5-1.0 mbar level and smoothing to lower pressures.  Furthermore, nadir retrievals cannot fully capture the oscillatory structure of Saturn's equatorial temperature profile, so low-latitude hydrocarbon distributions should be treated with caution. The colours correspond to different $L_s$, as labelled.}
\label{cxhy}
\end{center}
\end{figure}

We therefore focus this section on new results on ethane and acetylene spanning the duration of Cassini's mission, as shown in Fig. \ref{cxhy}.  Both \citet{09guerlet} and \citet{15sylvestre} utilised the enhanced vertical resolution of CIRS limb-sounding to measure the vertical distributions in the 0.01-5.0 mbar range, enabling a comparison of pre-equinox (2005-2008) and post-equinox (2010) observations, and both datasets are shown as squares in Fig. \ref{cxhy}.  These are compared to crude estimates of ethane and acetylene distributions from CIRS nadir observations at low spectral resolution (15 cm$^{-1}$) but high temporal resolution.  Snapshots of these observations were presented by \citet{13sinclair, 14sinclair}, but individual spectral inversions are sensitive to calibration defects and extreme degeneracies with the determination of stratospheric temperatures.  In Fig. \ref{cxhy} we adopt the technique of \citet{17fletcher_QPO} and \citet{18fletcher_poles}, fitting tensioned splines as a function of time at each latitude to remove significant outliers and extract general trends.  The caveat is that any short-term variability (on the order of months) would be smoothed over, and that the 2011-2013 period perturbed by Saturn's northern storm had to be removed from the time series.   Polar results were previously shown by \citet{18fletcher_poles}, and the figures are extended here to cover all latitudes.  We caution the reader that nadir observations have limited sensitivity to the 0.1-mbar level, and so do not show the same large-scale variability as the limb measurements.
\index{ethane}\index{acetylene}\index{CIRS}

\adjustfigure{100pt}

\begin{figure*}%
\begin{center}
\figurebox{6.0in}{}{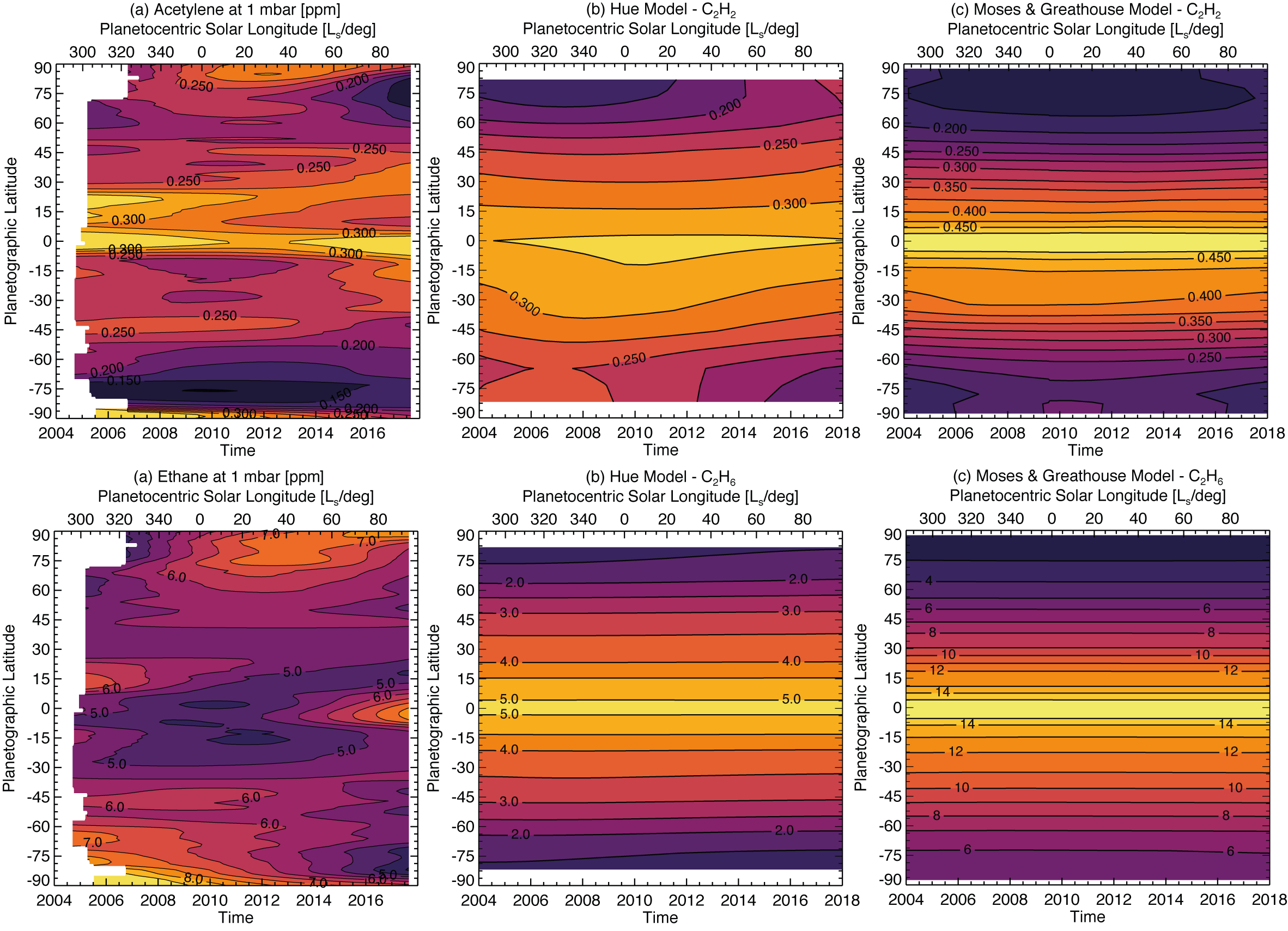}
\caption{Reconstructed mole fractions for ethane and acetylene at 1 mbar.  Left-hand column shows abundances from CIRS nadir observations, constructed by fitting tensioned splines through the time series of retrievals at each latitude as described by \citet{18fletcher_poles}.  Variability smaller than the difference between contour levels (i.e., smaller than 0.05 ppm for C$_2$H$_2$, or 1 ppm for C$_2$H$_6$) is likely due to uncertainties in the spectral inversion.   Central and right column show predicted abundances based on the photochemical modelling of \citet{15hue} and \citet{05moses_sat} over the 2004-2018 time span of the Cassini mission, highlighting the discrepancy between data and models.}
\label{cxhy_contour}
\end{center}
\end{figure*}

The nadir time series is also shown in Fig. \ref{cxhy_contour} for the 1-mbar altitude, and is compared to the seasonal predictions of \citet{15hue} and \citet{05moses_sat}, discussed below.  Several features of the latitudinal distributions stand out:  \begin{itemize}
    \item The C$_2$H$_2$ distribution at 0.1 and 1 mbar shows a strong decline from the equator to $\sim70^\circ$ latitude in both hemispheres, tracking the annual-average insolation, whereas the C$_2$H$_6$ is more latitudinally uniform with a rise towards high latitudes.  A similar situation occurs on Jupiter \citep{18hue}, and to date, there is no chemical explanation for the difference between these two species, which are expected to be coupled and to both show similar equator-to-pole gradients (Fig. \ref{cxhy_contour}).\index{acetylene}
    \item By summer solstice, both species show an equatorial maximum in abundance within $\pm10^\circ$ latitude, but the local equatorial peak in the C$_2$H$_6$ distribution developed only after 2012, possibly in association with Saturn's Equatorial Stratospheric Oscillation.  The C$_2$H$_2$ maximum appears relatively stable over the course of the dataset.  However, these nadir retrievals cannot resolve the vertical temperature structure associated with Saturn's equatorial oscillation \citep{08fouchet}, so the equatorial hydrocarbon results within $\pm10^\circ$ should be treated with caution. \index{equatorial oscillation}
    \item There are hints of complex temporal variability in the $10-25^\circ$ latitude ranges in both hemispheres, but more pronounced in the north.  The seasonally-reversing Hadley cell of \citet{12friedson} predicts subsidence of hydrocarbon-rich air near $25^\circ$ in the winter hemisphere, consistent with the local maximum in the $10-25^\circ$N region in Fig. \ref{cxhy} for ethane and acetylene. This winter subsidence appears to weaken at 1 mbar over time \citep[comparing $L_s=308^\circ$ to $L_s=134^\circ$ in][]{12friedson}, with diffusion helping to remove the local maximum in the later nadir and limb data by the spring \citep{15sylvestre}.  The effects of Saturn's ring shadow \citep{09guerlet}, and the downward propagation of the extra-tropical features of Saturn's Equatorial Stratospheric Oscillation \citep{17fletcher_QPO}, may also be playing a role at these latitudes. \index{equatorial oscillation} \index{Hadley cell} \index{ring shadow}
    \item Limited seasonal change is apparent at Saturn's mid-latitudes in Fig. \ref{cxhy}, in partial disagreement with earlier results \citep{13sinclair}.
    % , although a subtle hemispheric asymmetry exists with larger abundances in the north than in the south, suggesting northwards transport of hydrocarbon air during the southern summer and autumn. 
    \item Conversely, the largest changes are occurring within the Northern and Southern Polar Stratospheric Vortices \citep[NPSV and SPSV,][]{18fletcher_poles}, with declining abundances at the south pole (potentially due to upwelling of hydrocarbon-poor air in southern autumn) and increasing abundances at the north pole (potentially due to subsidence of hydrocarbon-rich air in northern spring), or with as-yet-undetermined auroral influences on polar chemistry. \index{polar stratospheric vortex}
\end{itemize}

Another important result from Cassini/CIRS was to show that seasonal variations of stratospheric composition are more muted than initially predicted.  Although the 1-mbar trends are expected to be small, one-dimensional photochemical models at lower pressures predicted seasonal abundance variations up to a factor of 10 at 0.01 mbar and high latitudes for C$_2$H$_2$ \citep{05moses_sat, 15hue}. Despite the limited temporal coverage of limb observations over one season ($\Delta L_s \sim 90^{\circ}$), such large-scale changes have not been observed. The absence of such strong contrasts may imply that meridional transport is an efficient process that redistributes the seasonally-produced hydrocarbons.  The discussion above hints at a complex balance between photochemistry and atmospheric circulation, discussed in the next section.

\subsection{Stratospheric Chemistry Modelling}
\label{chemmodel}

Proper modelling of the stratospheric physico-chemistry implies coupling a general circulation model that calculates the spatio-temporal distribution of several tens of compounds while simultaneously accounting for (photo-)chemical reactions. This is an especially computationally demanding task since the production of higher-order hydrocarbons, and the subsequent formation of hazes and aerosols, implies hundreds of chemical reactions \citep[e.g.,][for Jupiter]{00wong, 03wong}. Furthermore, uncertainties in the chemical reaction rates have to be considered, as they affect the accuracy of photochemical model predictions \citep{03dobrijevic}, particularly for the higher-order hydrocarbons. Reduced chemical networks allow computing the abundances of a smaller subset of compounds with great fidelity while limiting the computational cost \citep{11dobrijevic}. 
\index{stratospheric composition}\index{photochemistry}\index{aerosols}

\adjustfigure{100pt}

No models have yet succeeded at self-consistently accounting for that coupling, although different approaches have been taken in that direction. \citet{12friedson, 10greathouse, 10dowling, 14guerlet, 20spiga} developed general circulation models that include the effect of heating and cooling from the main stratospheric compounds but, as described in Section \ref{temp}, the spatial distributions of the main coolants (i.e., C$_2$H$_2$ and C$_2$H$_6$) are assumed to stay constant in time, counter to the changes observed in Fig. \ref{cxhy}.  Conversely, photochemical models generally use a pseudo-2D (altitude-latitude) approach, either by modelling a sum of 1D vertical columns at different latitudes \citep{05moses_sat}, or as a 2D-model without meridional transport \citep{15hue}, equivalent to a 1D approach.  Atmospheric circulation is neglected in these photochemical models for Saturn, but there have been attempts to couple photochemistry and transport for Jupiter \citep{18hue}. In our previous review \citep{18fletcher_book}, we presented a comparison between the photochemical predictions of \citet{05moses_sat} and the hydrocarbon abundances recorded in the CIRS nadir and limb observations.  In Fig. \ref{cxhy_contour} we therefore restrict the comparison to ethane and acetylene at 1 mbar, considering the model of \citet{16hue} which includes feedback between the temperature and abundance fields as explained in Section \ref{temp}.  The key conclusion from Fig. \ref{cxhy_contour} is that the hydrocarbon observations suggest a more complex behaviour than the models can currently account for.
\index{photochemistry}\index{circulation}
% The modification of the thermal field has a small feedback effect on the abundance distributions, which never exceeds 11\% and 9\% for C$_2$H$_2$ and C$_2$H$_6$.

One-dimensional fixed-season photochemical models have also been developed to explain the observed abundances of CH$_4$, C$_2$H$_2$, C$_2$H$_4$, C$_2$H$_6$, and C$_6$H$_6$ in the stratospheric homopause region of Saturn, as retrieved from the \textit{Cassini} UVIS occultations \citep{16koskinen}. These models focused on specific latitudes and seasons relevant to the UVIS occultations, and the results indicated  that neutral photochemistry alone was insufficient to explain the observed hydrocarbon abundances, particularly with respect to benzene.  An extra hydrocarbon production source is needed in the upper stratosphere, suggested to be ion-neutral chemistry initiated by either absorption of solar extreme ultraviolet radiation (which has a seasonal dependence), or by interaction of magnetospheric charged particles with the atmosphere in the auroral regions (which has a latitude dependence), or a combination of the two. The fact that the benzene mixing ratio is observed to be more than an order of magnitude greater at high northern latitudes in spring, in comparison with similar high southern latitudes in late autumn after this region entered the perpetual darkness of polar winter, has prompted \citet{16koskinen} to conclude that solar-driven ion chemistry plays an important role in C$_6$H$_6$ production --- auroral chemistry alone and subsequent transport away from the auroral regions is not the sole mechanism for producing and distributing benzene on Saturn. 
\index{UVIS}\index{occultation}\index{homopause}\index{benzene}
\index{ion chemistry}\index{auroral chemistry}

% In these models, the season and latitude have been held fixed to specifically compare with the UVIS results at 48.9$^{\circ}$ S latitude in April 2005 ($L_s$ = 303.5$^{\circ}$, southern summer), 74.3$^{\circ}$ N latitude in September 2012 ($L_s$ = 36.9$^{\circ}$, northern spring), and 72.7$^{\circ}$ S latitude in January 2015 ($L_s$ = 63.7$^{\circ}$, late southern autumn).  Based on comparisons of these models with the retrieved UVIS hydrocarbon profiles, \citet{16koskinen} determine that neutral photochemistry alone is insufficient to explain the observed hydrocarbon abundances, particularly with respect to benzene, such that an extra hydrocarbon production source is needed in the upper stratosphere. 

%The following paragraph is based on a DPS abstract, so you can delete it if that does not jibe with the rules for the book – but given that it took years to get the chapter published last time, there may be a full publication by the time the book is finalized
Further ion-neutral photochemical modelling by \citet{18moses_dps} confirmed the conclusions of \citet{16koskinen} in that solar-UV driven ion chemistry results in a factor of $<\sim10$ increase in the abundance of benzene and polycyclic aromatic hydrocarbons (PAHs) in comparison with models that consider neutral chemistry alone; however, the models also show that ion chemistry has less of an influence on the abundance of smaller C$_2$H$_x$ hydrocarbons.  \citet{18moses_dps} determined that solar-driven ion chemistry leads to a strong seasonal dependence of the benzene production rate, with more C$_6$H$_6$ being produced in summer than winter.  These models, which include solar chemistry but no auroral chemistry, underpredict the abundance of neutral hydrocarbon photochemical products at the high stratospheric altitudes probed by UVIS at high latitudes near the auroral regions.  As discussed by \citet{18moses_dps}, auroral ion chemistry could potentially make up some of this shortfall but is unlikely to fully resolve the model-data mismatch, suggesting that atmospheric dynamics could be affecting high-altitude species’ distributions on Saturn.  
\index{ion chemistry}

%The following two paragraph discusses modeling that has nothing to do with seasons, so you can decide whether to include it or not, but it won’t have a home in any other book chapter
Recent non-seasonally-dependent chemical models have confirmed the influence of exogenic material on Saturn's stratospheric chemistry.  Oxygen and other elements can be delivered via cometary impacts, interplanetary dust particles, or local satellite/ring sources, with H$_2$O, CO, and CO$_2$ becoming the dominant stable products from oxygen chemistry \citep{17moses}.  From models of dust-particle dynamics in the outer solar system, and models of dust ablation and subsequent stratospheric photochemistry in Saturn’s atmosphere, \citet{16poppe} and \citet{17moses} concluded that the influx of interplanetary dust particles was insufficient to explain the relatively large abundance of H$_2$O, CO, and CO$_2$ in Saturn’s stratosphere \citep[e.g.,][]{97feuchtgruber}.  Ever since active plumes were discovered on Enceladus by Cassini \citep{06dougherty,06porco}, this icy moon has been suggested as the dominant source for Saturn’s stratospheric water \citep[e.g.,][]{10guerlet,10cassidy,11hartogh,12fleshman}, and spatially resolved stratospheric H$_2$O observations by Herschel have validated this conclusion \citep{19cavalie}. Saturn’s stratospheric CO, on the other hand, may result from a large cometary impact that occurred several hundred years ago \citep{10cavalie,17moses}.   The large influx of ring vapour observed to be entering Saturn's upper atmosphere from \textit{in situ} measurements during the Grand Finale stage of the Cassini mission \citep{18waite, 20miller, 22serigano} is also expected to affect Saturn's stratospheric chemistry \citep{22moses}.  However, the expected chemical consequences were not seen by contemporaneous Cassini UVIS or CIRS observations, suggesting that the inferred large ring-vapour influx derived from a recent dynamical event in the rings \citep[see, e.g.,][]{19hedman}, or that most of the ring material enters the atmosphere as small dust particles that never actually ablate and therefore do not significantly affect stratospheric chemistry \citep[see][]{22moses}.  Chapters by Koskinen \textit{et al.} and Moore \textit{et al.} in this volume provide more detail on the effects of external material on Saturn’s thermosphere and ionosphere.
\index{exogenic species} \index{comet} \index{Enceladus torus} \index{oxygenated species} \index{water} \index{CO} \index{rings} \index{INMS}

In summary, Cassini has revealed slow seasonal changes in the distributions of C$_2$H$_x$ species over half a Saturnian year, as well as hemispheric asymmetries in other molecules, but these are yet to be fully explained by coupled neutral chemistry, ion chemistry, and atmospheric circulation models.  Continued tracking of spatio-temporal changes, alongside future model development, are needed to understand Saturn's seasonal variability.

%%%%%%%%%%%%%%%%%%%%%%%%%%%%%%%%%%%%%%%%%%%%%%%%%%%%%%%%%%%%%%%
%%%%%%%%%%%%%%%%%%%%%%%%%%%%%%%%%%%%%%%%%%%%%%%%%%%%%%%%%%%%%%%
%%%%%%%%%%%%%%%%%%%%%%%%%%%%%%%%%%%%%%%%%%%%%%%%%%%%%%%%%%%%%%%
\section{Aerosol Variability}
\label{aerosols}

At the time of the previous overview of our understanding of Saturn's aerosols by \citet{18fletcher_book}, there was a consensus that seasonal insolation changes induce hemispheric asymmetries in the tropospheric (and likely stratospheric) hazes, with higher opacity in the summer hemisphere and lower opacity and a bluer visual colour in the winter hemisphere.  Consistent asymmetries were also observed at near-IR wavelengths, with increased upper tropospheric opacity producing greater attenuation of Saturn's 5-$\mu$m emission in the summer hemisphere \citep{06baines_dps,11fletcher_vims}.  There was also a consensus that upper tropospheric aerosols reach to higher altitudes at the equator and stratospheric aerosols reach their greatest opacity in the polar regions.  As Saturn proceeded through its seasonal changes over the last 40 years, there has  been an evolution in instrumental capabilities and observational constraints that enables a more discriminating investigation of Saturn's aerosols, which we focus on in this improved but still incomplete assessment of seasonal aerosol variability.
\index{aerosols} \index{colour} \index{seasonal asymmetry} \index{ISS} \index{VIMS}

% But the latest reference to aerosol investigations in that previous review was published in 2013, and even those did not account for the full capabilities of the advanced observations available at the time.  Some of these capabilities have been used in more recent publications during the past seven years, and new results have been obtained from more recent observations through the end of the Cassini mission.  That is the focus of this summary of an 

\begin{figure*}%
\begin{center}
\figurebox{5.0in}{}{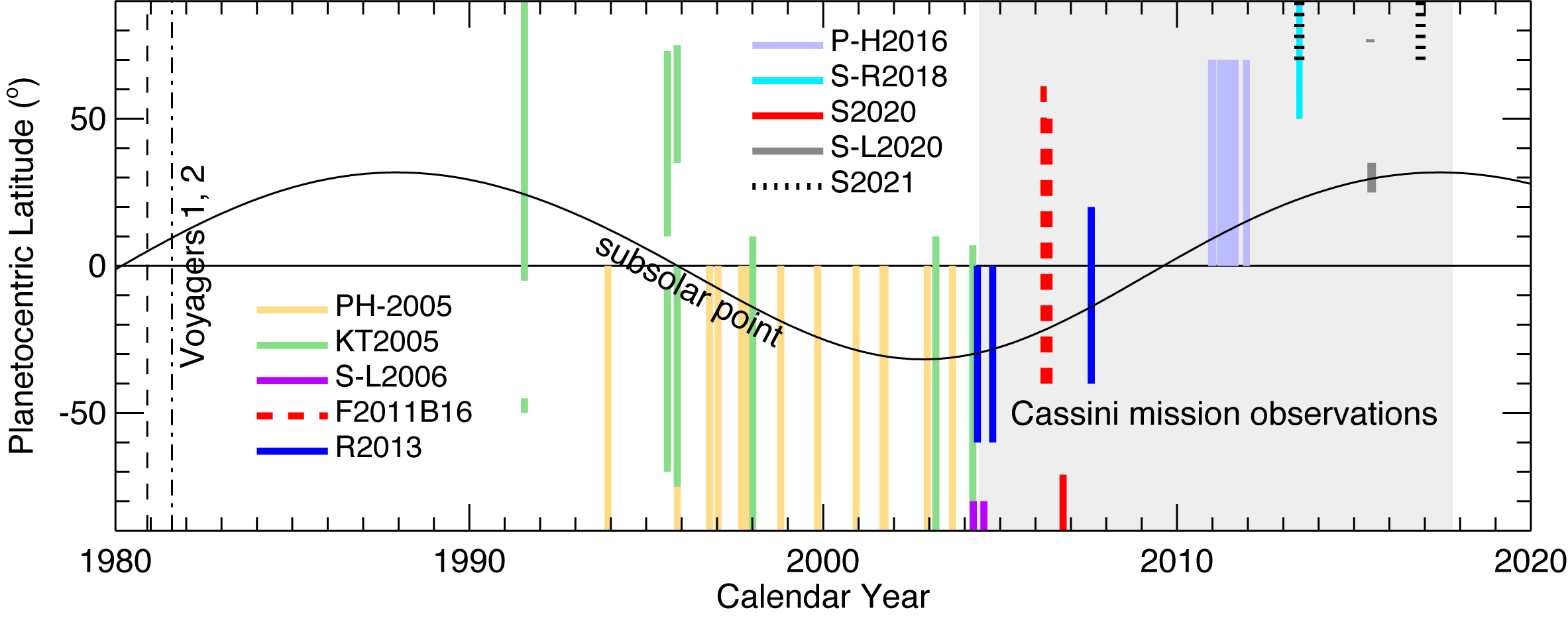}
\caption{Sub-solar latitude as a function of time, with temporal latitudinal sampling by published investigations of Saturn's aerosol distributions indicated by coloured bars for references indicated in the legend, which are \citet{05perez-hoyos}, \citet{05karkoschka}, \citet{06sanchez},  \citet{11fletcher_vims} and \citet{16barstow} (combined as by F2011B16), \citet{13roman}, \citet{16perez-hoyos},  \citet{18sanz-requena}, \citet{20sromovsky}, \citet{20sanchez_hex}, and \citet{21sromovsky}. Year labels are centred on tick marks placed at the beginning of the labelled year. These do not include storm-focused publications covering small local regions.}
\label{Fig:sampling}
\end{center}
\end{figure*}

\subsection{Cassini Observational Improvements in Context}  

Observations by Pioneer 11 in 1979, and Voyager 1 and 2 in 1980 and 1981, were limited in their spectral discrimination of vertical aerosol structure due to the use of broadband filters, yet they made important discoveries that are reviewed by \citet{09west} and more recently and more briefly by \citet{18fletcher_book}.  Here we focus on the analysis of observations by the Hubble Space Telescope (HST) between 1991 and 2004, and Cassini observations between 2004 and 2017.  The temporal and latitudinal coverage of a number of key investigations using these observations are displayed in Fig.\ \ref{Fig:sampling}.  
\index{Pioneer} \index{Voyager} \index{Hubble}

Relative to Voyager and Pioneer, HST observations provided more broadband and important narrow-band filters and sensitivity to methane band absorptions of different strengths, but were limited to low phase angles.  Capabilities to constrain vertical structure were improved significantly by Cassini. Its Imaging Science Subsystem (ISS) provided better sampling of methane band absorptions at CCD wavelengths, with greatly enhanced spatial resolution and phase angle coverage relative to HST observations.  Even more significant to aerosol investigations were observations by its Visual and Infrared Mapping Spectrometer (VIMS), an imaging spectrometer that covered wavelengths from 0.35 to 5.12 $\mu$m.  Along with this greater range of wavelengths, VIMS provided a wider variety of methane absorption strengths, and access to thermal emissions by Saturn, the combination of which facilitated the probing of deeper cloud structures, identification of cloud compositions, and constraining the spatial variability of phosphine gas (PH$_3$).  While the seasonal range of HST and Cassini observations offers great potential to characterise and understand seasonal variability in Saturn's aerosols, a comprehensive analysis including the full range of the spectral constraints offered by Cassini remains to be accomplished. Recent efforts have focused on storms \citep{16oliva,13sromovsky, 16sromovsky, 18sromovsky_storm} and polar regions \citep{06sanchez,18sanz-requena, 20sanchez_hex, 20sromovsky_poles, 20sromovsky, 21sromovsky}, with limited latitudinal and temporal sampling, and only one Cassini study \citep{21sromovsky} has combined both simultaneous coincident visual and near-IR constraints, using both reflected sunlight and thermal emission to constrain aerosol structure models in the north polar region. 
\index{VIMS} \index{ISS} \index{aerosols}
\adjustfigure{100pt}

\subsection{Aerosol Vertical Structure and Composition}  
\begin{figure*}%
\begin{center}
\figurebox{5.5in}{}{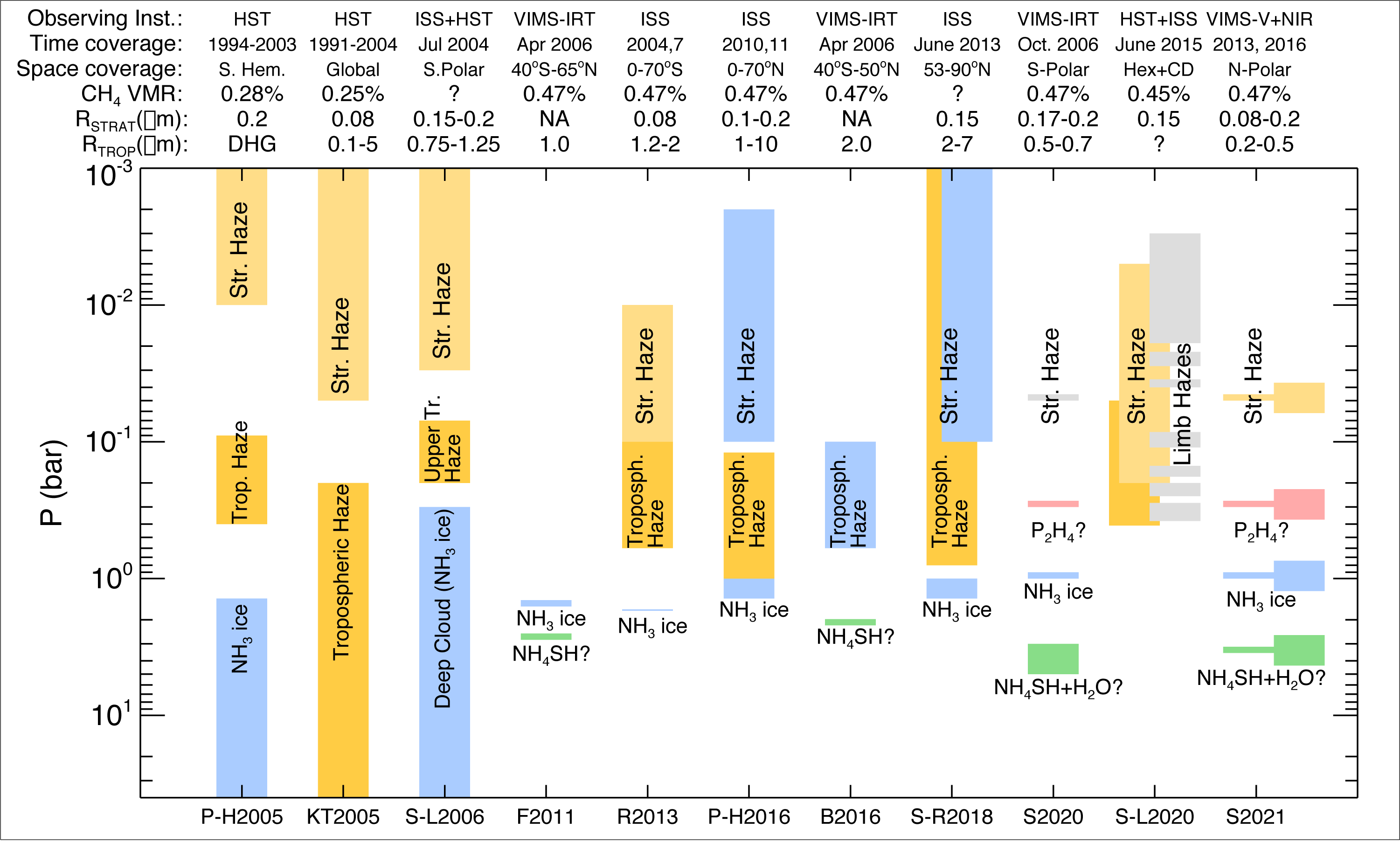}
\caption{Schematic models of vertical aerosol structures used to fit observations characterised in the legend table above each reference column. Note that successful models vary greatly in assumed vertical distributions. Reference labels are the same as in Fig.\ \ref{Fig:sampling}.  Note the multiple compact layers detected by limb observations analysed by \citet{20sanchez_hex}.  The yellow-orange colours mark layers with assumed or derived short wavelength absorption (the chromophore layers), with two shades used if different chromophore models are used in different layers. Stratospheric layers without shortwave absorption model fits are indicated in grey.  Blue denotes NH$_3$ ice particle layers, pink possible P$_2$H$_4$, and green possible NH$_4$SH or H$_2$O ice particles. ECCM models, e.g. \citet{05atreya}, predict an ammonia ice layer, and an NH$_4$SH layer with possible overlap of an even deeper water cloud. But ammonia gas depletion by the formation of the NH$_4$SH can alter the ammonia condensation level. The stratospheric and putative P$_2$H$_4$ layers are expected to originate from photochemistry.}
\label{Fig:structure}
\end{center}
\end{figure*}

Aerosol vertical structure models from the publications identified in Fig.\ \ref{Fig:sampling}, are crudely summarised in Fig.\ \ref{Fig:structure}.  Vertical structure results published prior to 2006 are skewed to some degree by the assumption of a methane volume mixing ratio of 0.28\%-0.25\%, which is well below the currently accepted value of 0.47\% \citep{09fletcher_ch4}. The visual and CCD observations were generally well matched by aerosol models that included two vertically extended particle layers (note that the use of semi-infinite layers in HST-based models is only a modelling convenience that is inconsistent with near-IR observations, because they would block Saturn's 5-$\mu$m thermal emission). The top layer in these models is a stratospheric haze of particles generally assumed to be spherical with radii of 0.1-0.2 $\mu$m, usually distributed between 10 mbar and 100 mbar. The second layer, termed tropospheric or upper tropospheric, consists of generally larger particles, with widely varying radii from 0.1 $\mu$m to 10 $\mu$m, depending on time, latitude, and model assumptions, with a vertical distribution also variable, but generally distributed between 100 mbar and 1 bar. This layer has an unknown composition, with diphosphine (P$_2$H$_4$) being the leading contender on photochemical grounds \citep{09fouchet}.  It is definitely not primarily ammonia ice because it lacks ammonia's 3-$\mu$m absorption signature, as inferred by \citet{97kerola} from ground-based observations, and by \citet{13sromovsky} from VIMS near-IR observations.  An alternative interpretation that the top layer might be primarily ammonia ice, but coated with other substances that obscure the ammonia spectral signatures, are less plausible because coatings do not hide those signatures very well unless they represent a significant fraction of the particle's mass \citep{89west}.
\index{aerosol vertical structure}
\index{ammonia ice}
\index{diphosphine}
\index{NH$_4$SH cloud}

\subsection{Revealing the Ammonia Layer} 

Many models illustrated in Fig.\ \ref{Fig:structure} include a deeper cloud layer assumed to be ammonia ice and placed near the pressure  predicted by equilibrium cloud condensation models (ECCM) \citep{05atreya}, with a nominal cloud base at 1.7 bars.  But this layer is difficult to constrain within the CCD spectral band, except for some discrete features of unusually high opacity \citep{13roman}.  To actually identify ammonia ice, which was initially expected to be the main constituent of the uppermost cloud layer on Saturn, requires detection of its strong 3-$\mu$m absorption feature and even better would be the additional detection of its much weaker 2-$\mu$m absorption feature. The VIMS near-IR spectrometer has the capability to make those detections, but the appearance of that spectral signature on Saturn has been extremely rare.  Until recently, ammonia ice has only been spectrally detected in the Great Storm of 2010--2011 \citep{13sromovsky}, and in Storm Alley \citep{18sromovsky_storm}, likely because vertical convection \citep[strongly suggested by associated lightning,][]{07dyudina, 13dyudina} had apparently elevated deeper condensates to the visible cloud level, or very near to that level in the Storm Alley case.  Additional examples were subsequently found in both polar regions \citep{18baines, 20sromovsky, 21sromovsky} where the overlying putative diphosphine layer was found to be sufficiently transparent that the ammonia ice layer is not entirely obscured by it. 
% In the undisturbed region ahead of the Great Storm of 2010--2011 \citet{13sromovsky} found this layer to have an optical depth of about 7 at 2 $\mu$m, which is ten or more times what is found within the polar regions. 
\index{equilibrium cloud condensation} \index{Great White Storm} \index{Storm Alley} \index{ammonia ice}

\adjustfigure{180pt}

\begin{figure*}%
\begin{center}
\figurebox{6.0in}{}{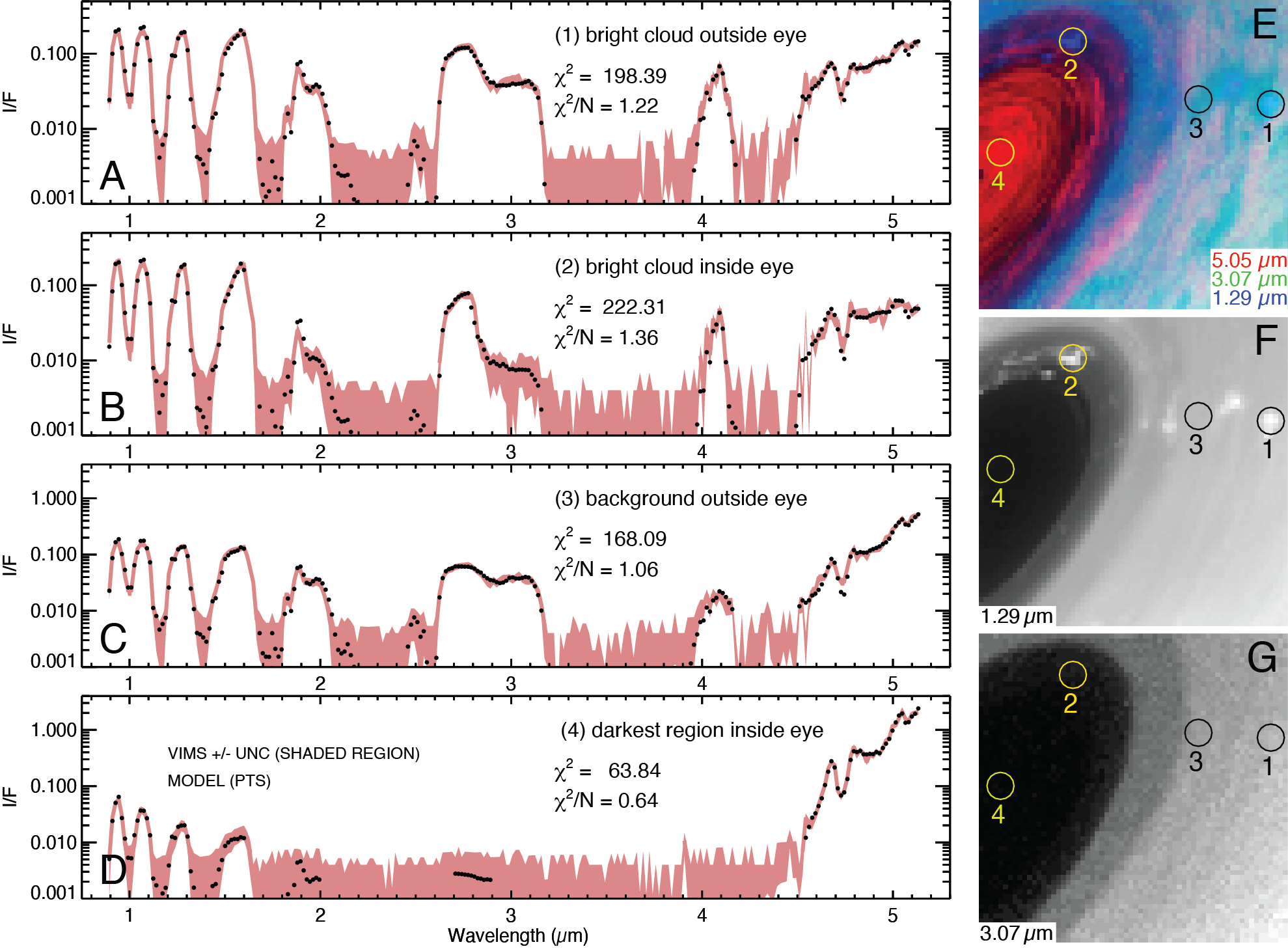}
\caption{VIMS near-IR observations of Saturn's north polar eye on 26 April 2017.  Discrete bright features at 1.29 and 5.05 $\mu$m exhibit strong absorption at 3.07 $\mu$m due to the presence of large ammonia ice particles. Adapted from \citet{18baines}.}
\label{Fig:eyespec}
\end{center}
\end{figure*}

The fact that the vertical location of this third layer is in a pressure-temperature range where NH$_3$ condensation is expected, and the fact that it displays the strong 3-$\mu$m absorption signature of ammonia, together argue that this layer is indeed primarily composed of NH$_3$ ice. One of the clearest examples of this detection is displayed in Fig.\ \ref{Fig:eyespec}, from an investigation of Saturn's north polar eye by \citet{18baines}. Note the ratios of reflectivity at 2.7 $\mu$m to that at 3.1 $\mu$m, which is about 2:1 for the background cloud spectrum in panel C, but more than 10:1 for the bright cloud spectrum in panel B. Note also that the ratio of reflectivity at 1.9 $\mu$m to that at 2 $\mu$m also is larger, but by smaller degrees, for clouds with increased 3-$\mu$m absorption.  The bright cloud features in panel F (a 1.29-$\mu$m image) are also bright at 4.09 $\mu$m, which makes them appear as magenta in the panel A colour composite because they lack reflectivity at 3.07 $\mu$m, shown in panel G.  The high I/F at 4--4.1 $\mu$m also suggests that NH$_4$SH is unlikely to be the main component of this 3-$\mu$m absorbing cloud.
\index{polar cyclone}

\subsection{Sounding the Deep Putative NH$_4$SH Cloud}

A layer of absorbing particulates in the 2-5 bar region is clearly required to modulate Saturn's 5-$\mu$m thermal emission. Plausible constituents are NH$_4$SH and water ice according to ECCM predictions \citep{05atreya}, but neither species provides an accurate spectral fit alone \citep{16barstow, 20sromovsky, 21sromovsky}. While NH$_4$SH provides better model fits, a mixed composition seems likely.  While Saturn's 5-$\mu$m emission seems to have a seasonal variation, indicated by its VIMS-observed hemispheric asymmetry in 2006 \citep{09west}, the deep layer itself provides no evidence of significant seasonal variation according to the modelling of VIMS night-side spectra by \citet{11fletcher_vims}.  They found that the best fits to the 4.6--5.1 $\mu$m spectra were obtained with a compact cloud centred in the 2.5-2.8 bar region, and a second aerosol layer at p$<$1.4 bars, with the latter being 1.5--2$\times$ more opaque in the summer hemisphere and mainly responsible for the observed asymmetry in 5-$\mu$m emission.  Whether the increased optical depths occurred in the putative diphosphine or ammonia layers could not be distinguished by the night-side observations, which lacked the constraints offered by reflected sunlight.  
\index{NH$_4$SH cloud} \index{VIMS} \index{aerosols}

In contrast to the apparent independent behaviours of upper tropospheric optical depths and the deep cloud pressures inferred by \citet{11fletcher_vims} from mid-latitude night-side spectra, \citet{20sromovsky} found in the south polar region that discrete bright features, produced mainly by increased optical depths in the ammonia layer, were also associated with the putative NH$_4$SH cloud tops being elevated to the 1.5--2 bar level. Averaging those elevated cloud pressures with the of 3--4 bars they inferred for the background clouds would put them closer to the zonal mean values of \citet{11fletcher_vims}.  However, there are other factors that might be affecting this pressure difference, including latitudinal differences in cloud structure, changes in VIMS calibration over time, and the use of different He/H$_2$ ratios in temperature extrapolations.

%\citet{Barstow2016} used a vertically diffuse upper layer
%extended between 0.1 and 0.6 bar, following \citet{13roman}, and
%tried to make use of limb darkening to further constrain cloud
%properties. The results were roughly similar to those of
%\citet{11fletcher_vims}, with slightly lower deep cloud pressures
%(near 2.3 bars generally), but significantly different gas mixing
%ratios, typically a factor of 2 higher than
%\citet{11fletcher_vims}. Both of these results lack the constraints
%provided by reflected sunlight, as well as the wide spectral range
%that can constrain aerosol particle size.  

\subsection{Aerosol Particle Shapes}  
\adjustfigure{60pt}

Many aerosol models have successfully matched observations using spherical particle scattering properties as a way to parameterise the wavelength dependence of optical depth and phase functions, even though none of the particle layers are expected to contain spherical particles.  Mie scattering is the most common model assumption for the stratospheric haze layer, but polarisation measurements in the polar region indicate that fractal aggregates might be the real physical form of particles in these regions \citep{14west, 18sayanagi_book}.  There is relatively uniform agreement that these stratospheric particles (when modelled as spheres) are in the range of 0.1-0.2 $\mu$m in radius, with the larger particles found after longer exposure to sunlight.  \citet{16perez-hoyos} used ISS observations over a wide range of phase angles to infer tropospheric layer phase functions, which they parameterised as double Henyey-Greenstein functions. They inferred that rather large (10-$\mu$m) particles may be present at low latitudes, with possibly 1-$\mu$m particles at intermediate latitudes.  Using models with spherical particle scattering yielded particle radius estimates ranging from tenths of a micron to 7 $\mu$m, depending on the observational constraints used, time and space sampling, and assumed model structure (see Fig.\ \ref{Fig:structure}).  It remains to be understood what interpretation problems exist for models using spherical particle phase functions, which can fit the observations well over at least a small range of phase angles, but do not match the phase functions inferred by \citet{16perez-hoyos} for the upper troposphere from observations taken over a wide range of phase angles.
\index{aerosol shapes}
\index{aerosol sizes}
\index{Mie scattering}

\subsection{Saturn's Chromophore}

\begin{figure}%
\begin{center}
\figurebox{3.5in}{}{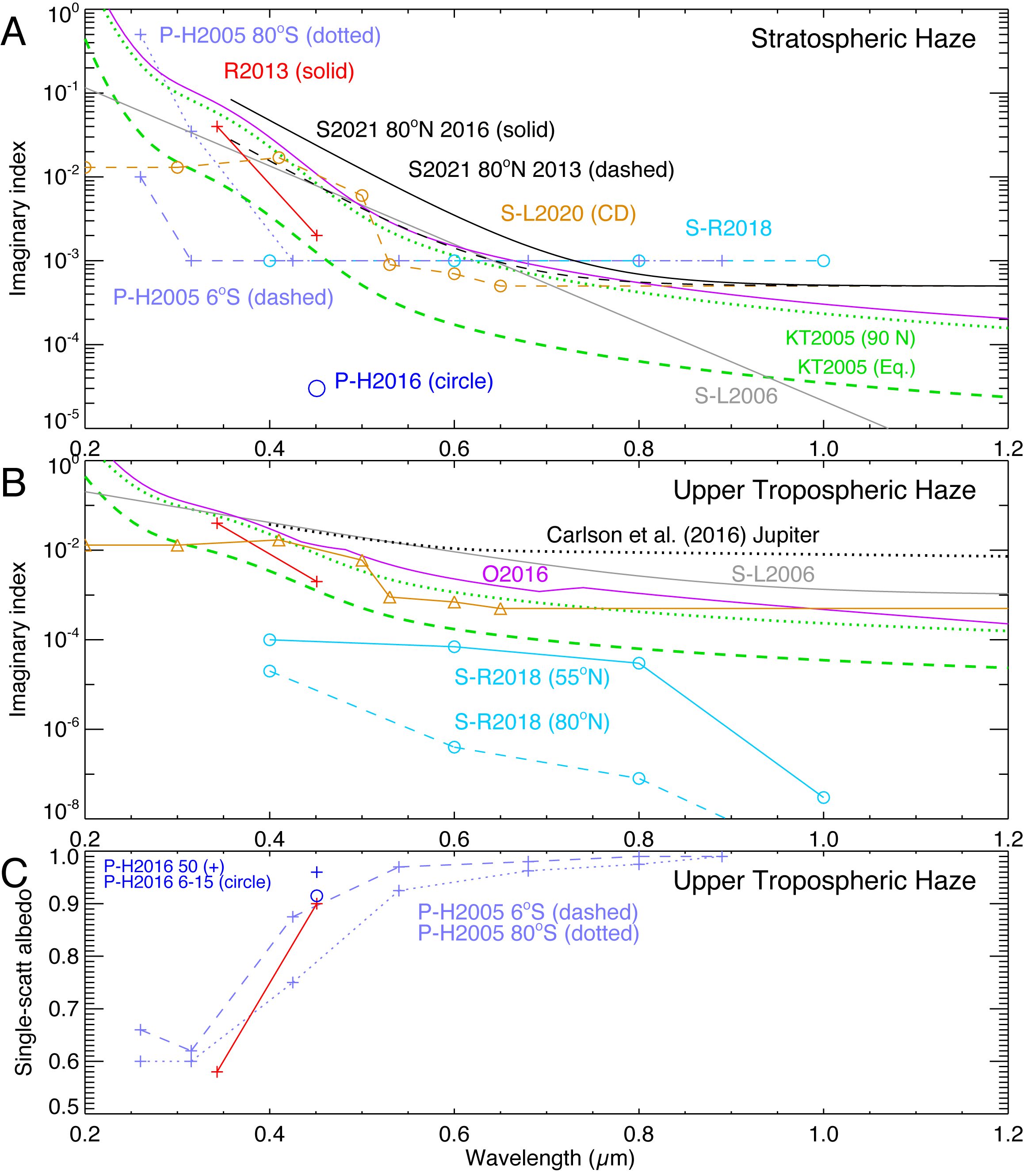}
\caption{Absorption properties of model chromophores for Saturn's stratosphere (A) and upper troposphere (B,C). Note the much slower falloff with wavelength of the \citet{16carlson} chromophore that has worked well for a variety of Jupiter observations.  For explanations of the abbreviations, see Fig. \ref{Fig:sampling}.}
\label{Fig:chrom}
\end{center}
\end{figure}

Seasonal aerosol variability is perhaps most directly observable in Saturn's visible colours, as revealed in Fig. \ref{saturn_montage}.  However, modellers disagree regarding the vertical location of aerosols responsible for Saturn's short-wavelength absorption, as indicated by the variety of yellow-orange layers in Fig.\ \ref{Fig:structure}. The weight of evidence seems to favour short-wavelength absorption at least being present in the stratosphere, as that has been detected in short-wavelength observations by Cassini UVIS imaging near 180 nm \citep{19pryor}, as well as in UV HST observations \citep{05karkoschka}.  There is also quite a variety in derived magnitudes of absorption, as illustrated in Fig.\ \ref{Fig:chrom}, which displays imaginary index versus wavelength derived for stratospheric and upper tropospheric layers (assuming spherical particles), and single-scattering albedo results for the upper troposphere (the only measure of absorption for double Henyey-Greenstein phase function models).   \citet{05karkoschka} found a large latitudinal variation in the magnitude of the imaginary index but not in its shape. \citet{21sromovsky} also found little spectral shape variation in the north polar hexagon region but a significant increase in amplitude between 2013 and 2016.  There is a surprisingly large unexplained variability among the many results displayed in Fig.\ \ref{Fig:chrom}, which might be due to different model assumptions or spatial and/or temporal variations that have yet to be fully characterised.
\index{colour} \index{chromophore} \index{UVIS} \index{ISS}

\subsection{Vertical Extent of Aerosol Layers}  

There is also a lack of clear constraint on the vertical distribution of particles within each compositional layer.  Early models constrained by the CCD spectral range were successful with broad vertical distributions.  But multiple compact layers, even in the stratosphere and upper troposphere, can also match observations, as illustrated by \citet{20sromovsky,20sromovsky_poles, 21sromovsky}. Evidence in favour of compact layers, at least in the south polar region, are evident from the presence of antishadows \citep{20sromovsky_poles} because physically thin layers are required to produce them.  Evidence for compact haze layers in the north polar region is their detection in 2015 ISS limb observations near the hexagon boundary (illustrated in Fig. \ref{Fig:structure}) and analysed by \citet{20sanchez_hex}, although it is not clear which of these layers have significant vertical optical depths (the fourth from the bottom shown in Fig.\ \ref{Fig:structure} appears to be the most prominent of the stratospheric layers).  There have also been shadows observed in the north polar region \citep{18ingersoll}, but these seem to be more subtle for reasons that remain to be understood.  \citet{21sromovsky} also showed that slightly better fits of north polar cloud structure can be obtained in some cases with thicker sheet clouds in which the top pressures are 60\% of the bottom pressures (vs. the 90\% they used generally).  Thicker layers of the same optical depths would also reduce shadow contrast.  More globally, compact layers also seem to be preferred by \citet{11fletcher_vims} for both deep and upper tropospheric layers, and at least for the deep layer by \citet{16barstow}, although these inferences lacked the additional constraints offered by observations in reflected sunlight.
\index{aerosol vertical structure} \index{polar cyclone}

%\adjustfigure{40pt}

\begin{figure*}%
\begin{center}
\figurebox{6.0in}{}{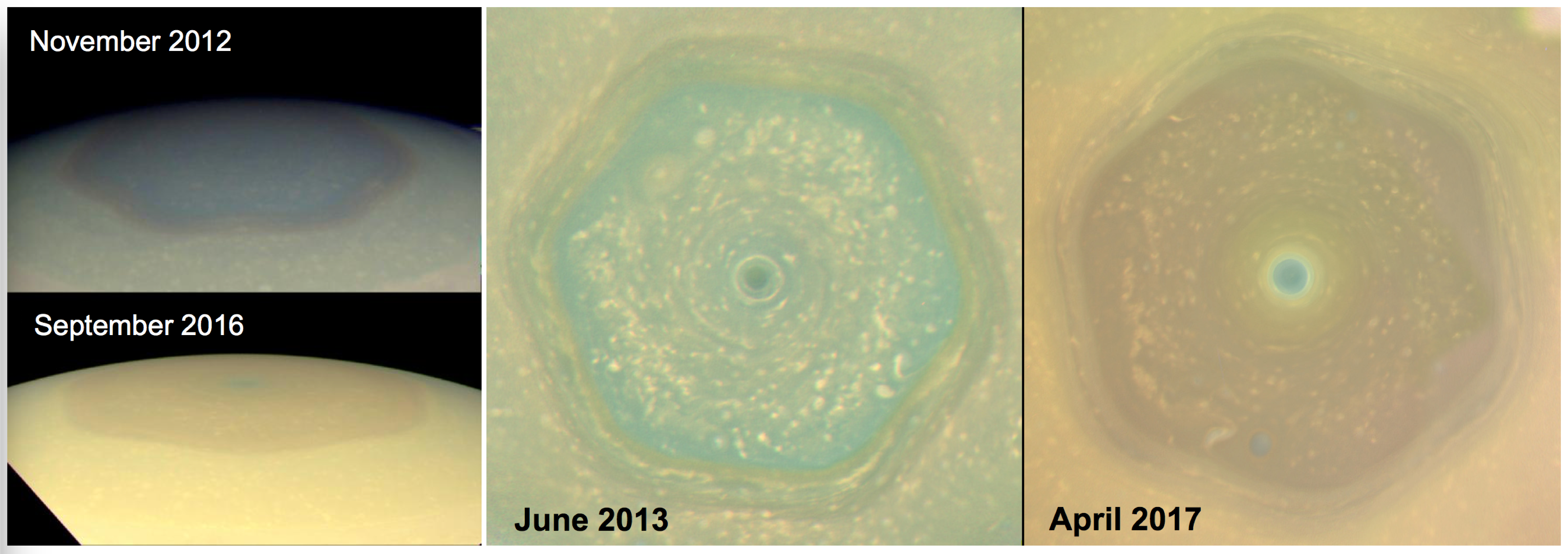}
\caption{Cassini ISS colour composite images documenting colour changes in Saturn's north polar region as it approached northern summer solstice. From \citet{21sromovsky}.}
\label{Fig:northcolor}
\end{center}
\end{figure*}

\subsection{Seasonal Changes in the North Polar Region}

\begin{figure*}%
\begin{center}
\figurebox{6.0in}{}{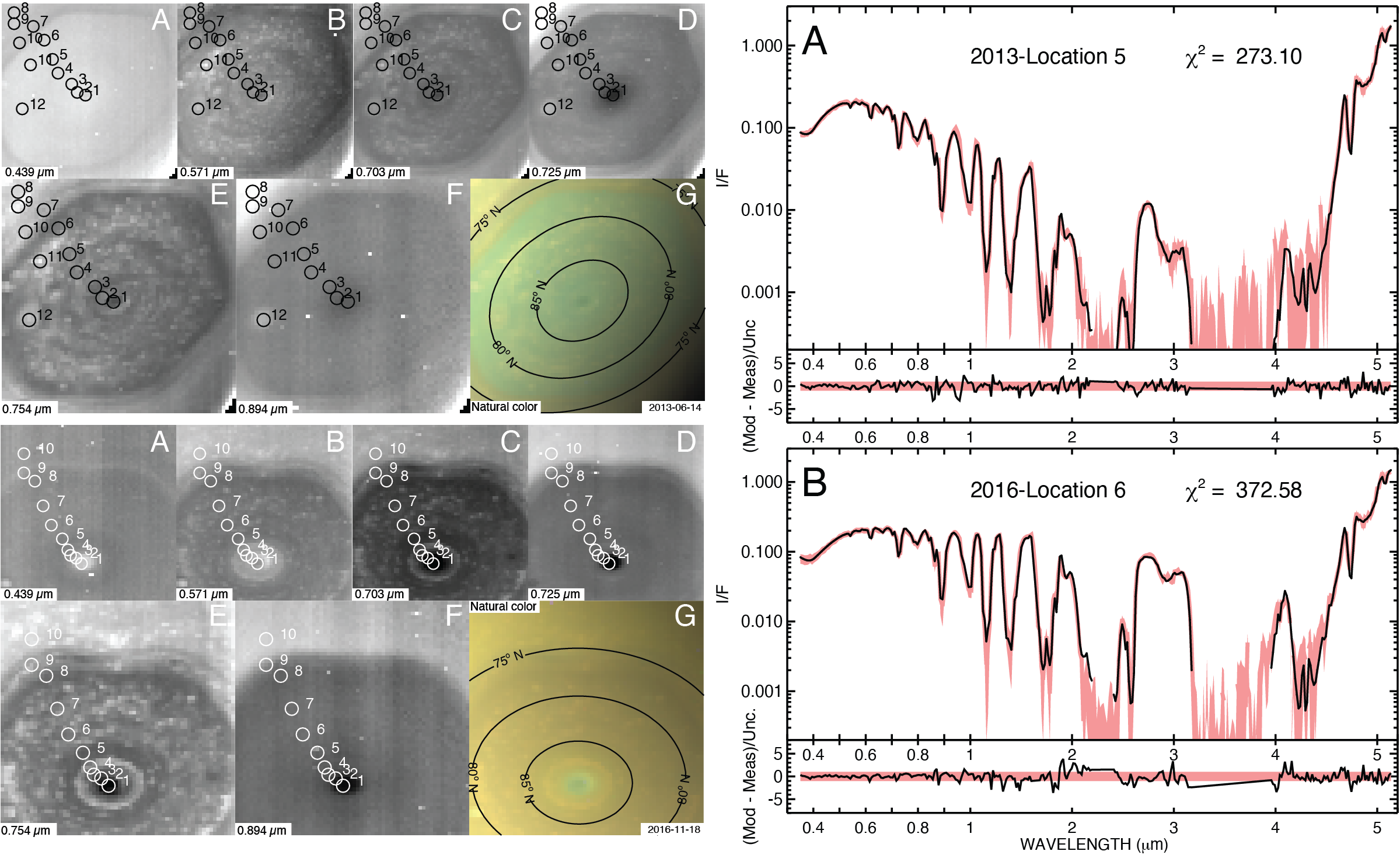}
\caption{Left: sample VIMS visual image planes and computed true colour image for June 2013 (top) and for November 2016 (bottom). Right: sample mid-hexagon spectra (coloured band width indicating uncertainty) from corresponding years and best-fit model spectra (solid black), with fit-difference/uncertainty plotted under each spectrum. From \citet{21sromovsky}.}
\label{Fig:northobs}
\end{center}
\end{figure*}

Cassini ISS images between 2012 and 2017 (Fig. \ref{Fig:northcolor}) document a significant change in the colour of the north polar region, from an initial blue/green colour throughout the interior of the hexagon, to a gold colour between the eye and the outer boundary of the hexagon (near 75$^\circ$N), but with little change seen within the eye (88.2$^\circ$ --90$^\circ$N).  VIMS observations in 2013 and 2016 (Fig.\ \ref{Fig:northobs}) helped to understand that evolution.   Visual and Near-IR observations in combination (right panels of Fig.\ \ref{Fig:northobs}), show that the changes in visual colour are accompanied by much larger spectral changes at near-IR wavelengths.  \citet{21sromovsky} employed both scattered sunlight and 5-$\mu$m thermal emission to constrain their four-layer vertical structure model (see Fig.\ \ref{Fig:structure}).  They found that the colour change could be entirely explained by changes in the stratospheric haze, which included a 2$\times$ increase in particle radius, a 5$\times$ increase in optical depth, and a 4$\times$ increase in peak short-wavelength absorption, without any significant change in the shape of the absorption spectrum.  They also found that the large increases at near-IR continuum wavelengths were mainly a result of large changes in the putative diphosphine layer (centred at 250-285 mbar), including a 2$\times$ increase in particle radius, and a 5$\times$ increase in optical depth. 
%(see Fig.\ \ref{Fig:northfits} for specific values). 
The ammonia layer changes were not quite as large, but important in explaining the changes seen in the 3-$\mu$m region.  
%The ammonia layer, between the discrete features, did not evolve much over this time %period (except for a central pressure decrease from 900 to 570 mbar), and in the eye very %low optical depths persisted. 
Very little change was seen within the eye, either at visible or near-IR wavelengths, with low optical depths persisting.  Apparently the increased sunlight that seemed to cause substantial change in the stratospheric haze throughout most of the hexagon interior, had little impact within the eye, which has a sharp boundary perhaps due to some dynamical transition that is also marked by a downwelling that would inhibit formation of aerosol layers. This is consistent with the thermal-IR inference of downwelling within the north polar cyclone by \citet{08fletcher_poles}. The blue/green to gold transition is also seen with lower contrast over a wide latitudinal range, in which the winter mid latitudes attain a slightly blueish cast, which fades to a more gold/yellow colour on exposure to sunlight (illustrated by Fig. \ref{saturn_montage}).
\index{chromophore} \index{colour} \index{diphoshine} \index{polar cyclone} \index{hexagon} \index{seasonal asymmetry}

\citet{21sromovsky} also indicate the first potential temporal changes in phosphine, previously alluded to in Section \ref{chem}.  Parameterising the vertical profile of PH$_3$ as a uniform mixing ratio up to a transition pressure, above which the mixing ratio decreased with altitude, they found that in a region between outer edge of the eye and the outer edge of the hexagon, the transition pressure was near to the pressure of the putative diphosphine cloud in 2016,  but was much deeper in 2013,
%They found that the transition pressure for the vertical PH$_3$ profile (from a profile %decreasing with altitude at low pressures above the transition level, or `breakpoint,' to %one that is uniform with depth at higher pressures) that
% outside of the eye but interior to the hexagon
 when that cloud had a much lower optical depth.  Within the eye itself, the transition point was found to be deeper than the surroundings in both 2013 and 2016, while the deep abundance of phosphine was found to be $\sim$30\% higher inside the eye than outside.  At higher altitudes in the upper troposphere, the combined effect of these parameters was to make the upper-tropospheric PH$_3$ abundance appear to be lower inside the eye, consistent with thermal-infrared studies \citep{08fletcher_poles}.
\index{phosphine} \index{diphosphine}

\adjustfigure{50pt}

\subsection{Summary of Aerosol Variability}

Much has been learned about the latitudinal distribution and seasonal forcing of Saturn's aerosol layers, as well as their behaviour in localised storm regions, and some progress has been made in identifying aerosol composition.  It is now reasonably clear that there are four compositionally distinct aerosol layers accessible to observations at UV to thermal infrared wavelengths.  Underneath a stratospheric UV-absorbing haze is a putative diphosphine layer that is effective at obscuring an underlying ammonia ice layer, except in polar regions, where the putative P$_2$H$_4$ layer is relatively translucent, and where Cassini/VIMS observations have made a strong case for the presence of ammonia ice, previously identified only in storm systems.  The deeper putative NH$_4$SH layer expected from ECCM models seems to be present but in combination with some other absorber. The tropospheric aerosols located at pressures less than 1.4 bar appear to undergo a seasonal variation, inferred from a hemispheric asymmetry in their obscuration of Saturn's 5-$\mu$m emission, though we still do not know whether that variation is in the putative diphosphine or ammonia ice layers.  This variation may be partially responsible for the temperature perturbation discussed in Section \ref{temp} and Fig. \ref{knee}, which is a radiative effect associated with these seasonally-variable aerosol layers. 
\index{aerosols}

The seasonal colour changes in the north polar region are now better understood.  Except for the eye of the north polar cyclone, the stratospheric, putative diphosphine, and ammonia ice layers all became significantly optically thicker as they were exposed to increasing sunlight between 2012 and 2017.  This caused a colour change from blue/green to gold, which is at least superficially similar to the observed hemispheric colour changes and those changes inferred from Hubble observations \citep{05karkoschka}.  Nevertheless, we still have much to learn about the global-scale aerosol variability away from the polar domain. Many Cassini datasets have only just begun to be exploited, including the full spectral range of VIMS observations over wider spans of latitude and time, in addition to the ISS imaging and polarisation measurements.  The connection between physical reality and parameterised models is currently somewhat uncertain, and while more extensive analysis of existing observations will help greatly, they will also need the support of laboratory investigations (a key priority is measuring the optical properties of diphosphine), photochemical and microphysical modeling, and ultimately suitable \textit{in situ} observations.
\index{diphosphine} \index{ammonia ice} \index{chromophore}

%%%%%%%%%%%%%%%%%%%%%%%%%%%%%%%%%%%%%%%%%%%%%
%%%%%%%%%%%%%%%%%%%%%%%%%%%%%%%%%%%%%%%%%%%%%
%%%%%%%%%%%%%%%%%%%%%%%%%%%%%%%%%%%%%%%%%%%%%
\section{Conclusions and Outstanding Questions}
\label{conclude}

Throughout Cassini's 13-year mission, its remote sensing experiments observed seasonal asymmetries in temperatures, atmospheric composition, and aerosols, and had begun to track how these asymmetries evolved as a function of time.  As was made clear in the discussion of radiative climate models (Section \ref{temp}), photochemical models (Section \ref{chemmodel}), and general circulation models, the seasonal trends are driven by insolation (to first order), but can be radically altered by the influence of dynamics, which in turn responds to contrasts in temperature.  The circulation of the upper troposphere and stratosphere is likely to vary on seasonal timescales --- from the seasonally-reversing Hadley cell at the equator with its interhemispheric transport \citep{09guerlet, 12friedson, 22bardet}, to the formation and dissipation of stratospheric polar vortices \citep{08fletcher_poles, 18fletcher_poles}, to the slow descent of temperature, wind (and potentially compositional) anomalies associated with Saturn's Equatorial Stratospheric Oscillation \citep{08fouchet, 08orton_qxo, 17fletcher_QPO, 18guerlet, 21bardet}.  
\index{Hadley cell} \index{equatorial oscillation} \index{polar stratospheric vortex}

\adjustfigure{40pt}

To understand how reproducible these dynamic perturbations are from Saturn-year to Saturn-year requires multiple decades of observations, exceeding the orbital period of Saturn (29.5 years).  Intriguingly, observations of the equatorial region one Saturnian year apart, from Voyager to Cassini \citep{14sinclair, 13li, 16fletcher, 22blake}, have suggested that the equatorial oscillation was not in the same phase (and therefore may not be semi-annual after all).  In 2011-2013 the oscillation was substantially disrupted by the mid-latitude `stratospheric beacon' that was produced by the Great White Storm \citep{12fletcher, 17fletcher_QPO, 18guerlet, 18sanchez_storm}, and the tropospheric aftermath of that storm (in terms of temperatures and aerosols) was apparent right until the end of the Cassini time series \citep{14achterberg, 16fletcher}, raising the question of whether sporadic storm eruptions could be a driver of interannual variability.
\index{Great White Storm} \index{interannual variability}

\begin{figure}%
\begin{center}
\figurebox{3.2in}{}{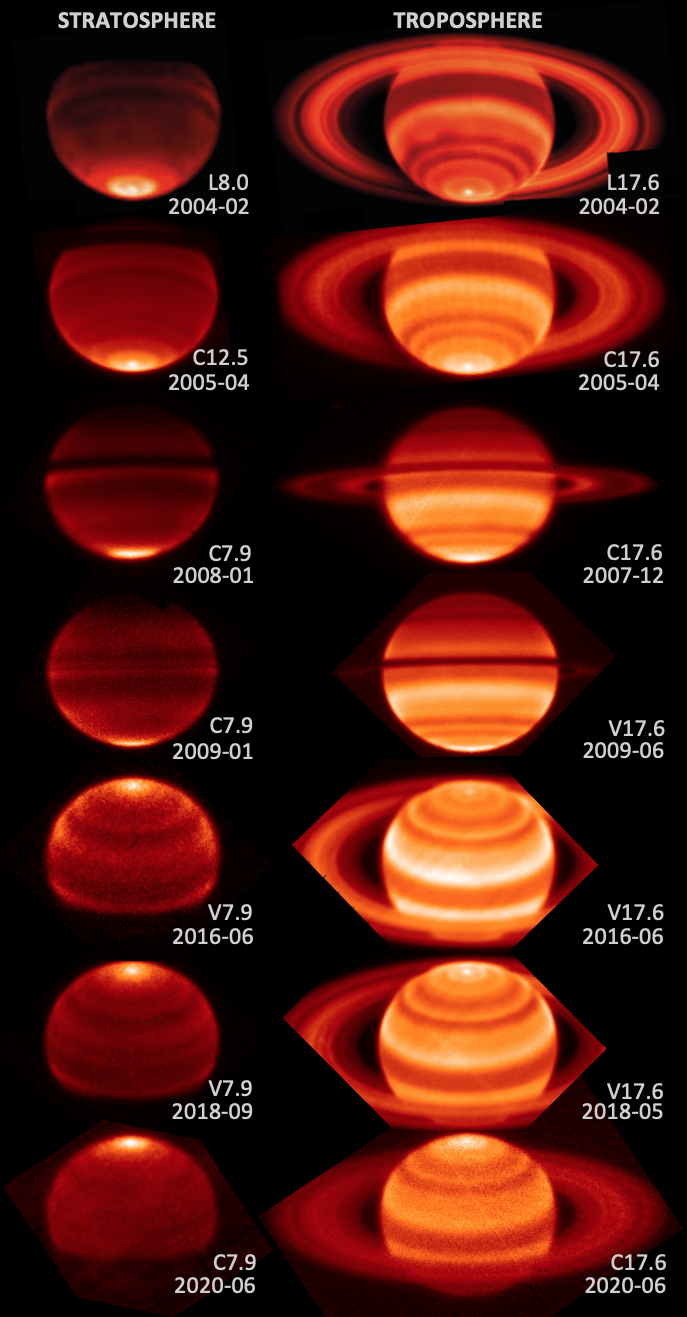}
\caption{Thermal images of Saturn from 2004 to 2020 adapted from \citet{22blake}, as observed by three 8m+ ground-based observatories:  L for Keck/LWS observations in 2004 \citep{05orton}, C for Subaru/COMICS \citep{09fletcher_imaging}, and V for VLT/VISIR.  Stratospheric images were typically acquired at 7.9-8.0 $\mu$m sensing stratospheric methane, with the exception of the 2005 observations at 12.5 $\mu$m sensing stratospheric ethane.  Tropospheric images were acquired at 17.6 $\mu$m.}
\label{thermalimages}
\end{center}
\end{figure}

\begin{figure}%
\begin{center}
\figurebox{3in}{}{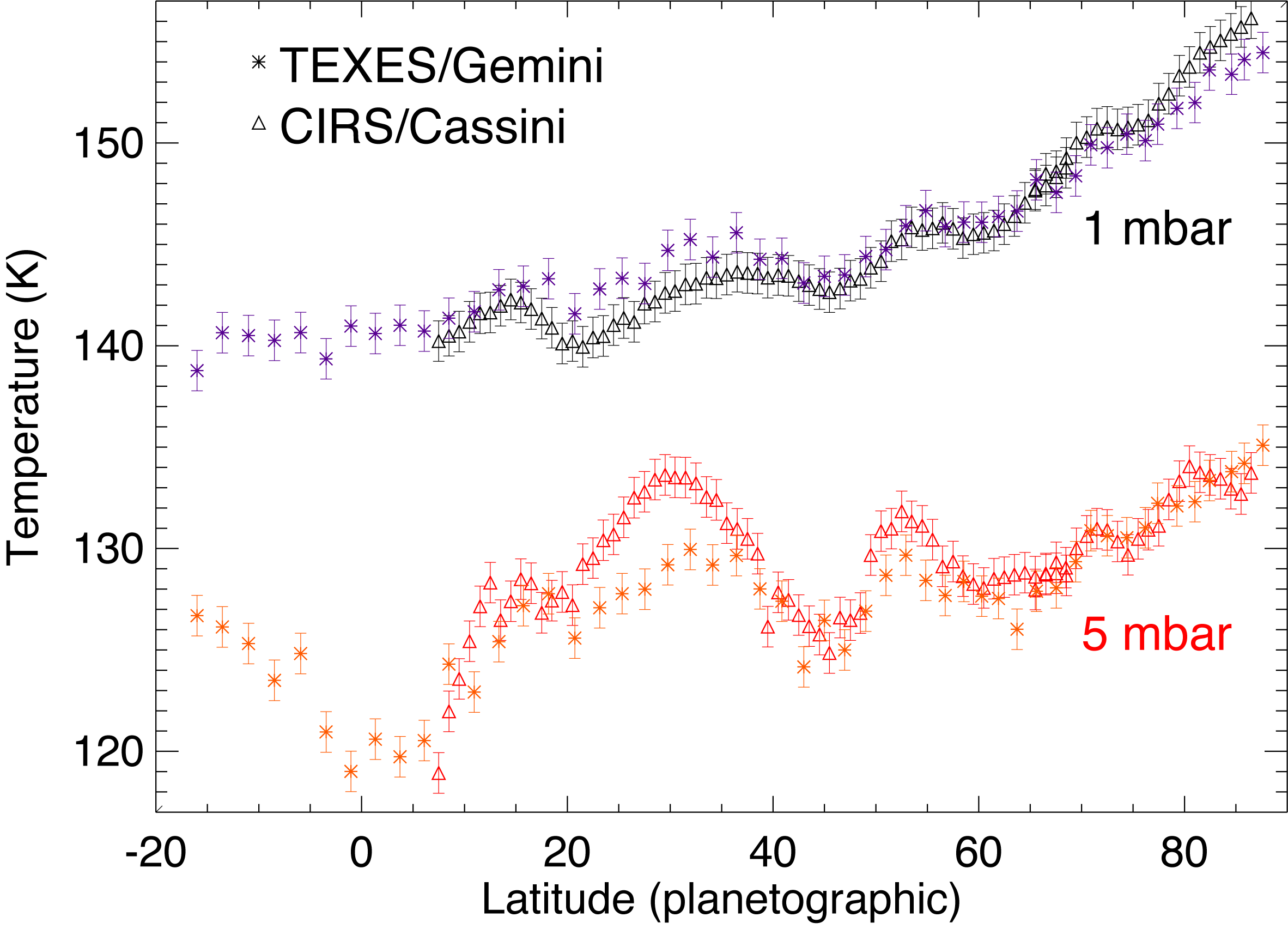}
\caption{Comparison of stratospheric temperature contrasts in Saturn's northern spring between Cassini/CIRS nadir observations in January-February 2017 (triangles) and Gemini/TEXES ground-based observations in March 2017 (crosses) at 1 mbar (top) and 5 mbar (bottom) \citep{17guerlet_dps}.}
\label{TEXES-CIRS}
\end{center}
\end{figure}

Efforts are now underway to understand interannual variability by comparing the Cassini record of temperatures, hydrocarbons, and aerosols (2004-2017) to ground-based observations since the 1980s and into the 2020s \citep{22blake}.  Fig. \ref{thermalimages} shows a montage of mid-infrared imaging of Saturn from 2004 to 2020, acquired by Keck \citep{05orton}, Subaru \citep{09fletcher_imaging} and the Very Large Telescope \citep{17fletcher_QPO} in narrow-band filters sensing stratospheric methane emission (7.9 $\mu$m) and the tropospheric H$_2$-He continuum (17.6 $\mu$m).  These provide the means to crudely determine the vertical temperature structure across the Earth-facing hemisphere, tracking the evolution of the banded structure and seasonal asymmetries as a function of time \citep{22blake}.  These are supported by spatially-resolved spectral maps from the TEXES instrument mounted on Gemini \citep{17guerlet_dps}, and Fig. \ref{TEXES-CIRS} demonstrates the accuracy with which Saturn's meridional stratospheric temperature gradient can be derived from these spectra, in comparison with CIRS observations taken nearby in time.  Both the imaging and spectroscopic datasets reveal that we can now do from the ground what was only previously possible from space, at least for the Earth-facing hemisphere.  Visible and infrared observations of this type will support forthcoming observations from the James Webb Space Telescope \citep{16norwood}, which launched in 2021.
\index{interannual variability} \index{ground-based observing} \index{James Webb Space Telescope}

Observations since 2017 extend the legacy of Cassini's exploration into a completely new season --- northern summer --- for the first time. Unfortunately, the Earth-based vantage point will only ever be able to access the sunlit hemisphere, so the winter hemisphere will be obscured in darkness for a generation, until some future orbital mission once again reaches Saturn.  Until then, scientists will monitor the evolution of temperature, composition, and aerosols through northern autumnal equinox (2025, $L_s=180^\circ$) and on to southern summer solstice (2032, $L_s=270^\circ$), completing the seasonal study through to the ``one-Saturn-year anniversary'' of Cassini's arrival ($L_s=293^\circ$) in 2034.  Such studies may allow us to address the following open questions:
\index{solstice} \index{equinox}

\begin{itemize}
\item How do aerosols, transport of photochemical products, and meridional circulation contribute to Saturn's seasonal temperature gradients?
\item What processes drive belt/zone contrasts in temperature, aerosols, and composition, and what causes the decay of the zonal winds with altitude?
\item What processes govern the dynamical evolution of Saturn's equatorial oscillation? It is mainly seasonally forced (like the semi-annual oscillation on Earth) or is it also shaped by variations in meteorological activity?
\item What is the deep distribution of gases below Saturn's visible cloud layers, and how do the belts and zones (and their associated wind/temperature gradients) vary with depth?
\item What controls the formation, colouration, and vertical distribution of aerosols, and what are these aerosols made of?
\item How do Saturn's auroras influence polar chemistry, aerosol formation, and heating?
\item How is the circulation of the ionosphere and thermosphere connected to seasonal contrasts in Saturn's stratosphere?
\item What determines the difference in meridional trends between C$_2$H$_2$ and C$_2$H$_6$, and how do photochemical species vary with latitude/time at high stratospheric altitudes?
\end{itemize}

Seasonal exploration requires the continued development of new modelling tools.  Since our last review \citep{18fletcher_book}, we have seen the combination of stratospheric photochemistry and radiative climate models \citep{16hue} and the combination of radiative calculations with general circulation models \citep{12friedson, 14guerlet,  20spiga}, but no attempts to date to combine all three into a self-consistent scheme.  Tropospheric photochemical modelling (particularly in relation to PH$_3$ and, to a lesser extent, NH$_3$) remains in its infancy, with the relationship to the seasonal variability of upper-tropospheric and stratospheric hazes remaining to be explored.  Microphysical modelling of cloud formation and precipitation, beyond the simplistic assumptions of thermochemical equilibrium cloud condensation, needs to be incorporated into circulation and chemistry models as active ingredients to understand the complex feedback between dynamics, chemistry, and cloud formation.  Furthermore, the remote sensing measurements require a greater degree of consistency than is typically provided by individual studies in restricted wavelength ranges.  Progress has been made in combining visible and near-infrared spectroscopy \citep[from ISS and VIMS,][]{21sromovsky}, but this needs to be extended to thermal wavelengths (CIRS) and ultraviolet wavelengths (UVIS).  Microwave measurements of Saturn provide insights into the connection between the upper troposphere and the deeper layers below the clouds \citep{13janssen}, but they remain underutilised, and could be continued from ground-based facilities like ALMA and the VLA.  Our characterisation of this seasonal giant, using both existing data and future measurements, will entail a broad combination of techniques and there are no shortages of opportunities for future research.
\index{radiative-climate model} \index{general circulation model} \index{circulation} \index{equilibrium cloud condensation} \index{cloud microphysics}

\section*{Acknowledgements}  

Fletcher was supported by a Royal Society Research Fellowship and European Research Council Consolidator Grant (under the European Union's Horizon 2020 research and innovation programme, grant agreement No 723890) at the University of Leicester.  Moses was supported by NASA Solar System Workings grant 80NSSC20K0462.  Sromovsky was supported by NASA Cassini Data Analysis Program grants 80NSSC18K0966 and 80NSSC22K0339.  The authors thank A.J. Friedson and M. Sylvestre for providing their data for the figures in this review.

\bibliography{references}\label{refs}
\bibliographystyle{cambridgeauthordate-Leigh}

\backmatter

%\begin{table}
%\caption{Missions and Investigations Relevant to Mars Surface Science: 1988--2007}{\tabcolsep5.5pt%
%\begin{tabular}{@{}lllll@{}}
%\toprule%
%Time, t(s)&$r_{N1}$ (cm)&$r_{N2}$ (cm)&$r_{N3}$ (cm)&$r_{N4}$ (cm)\\
%\hline
%10 & 8.2  & 8.6   & 8.5   & 8.0 \\\hline
%15 & 8.1  & 8.1   & 8.1   & 8.5   \\\hline
%30 & 8.5  & 8.5   & 9.1   & 9.3 \\\hline
%45 & 9.2  & 9.2   & 9.2   & 9.5   \\\hline
%60 & 9.5  & 9.6   & 9.8   & 9.8   \\\hline
%90 & 9.8  & 1.0   & 1.0   & 1.3    \\\botrule
%\end{tabular}}
%\begin{tabnote}
%Notes: Investigations discussed in this book, CRISM (Compact Reconnaissance Imaging Spectrometer for Mars), CTX (Context
%Camera); GRS (Gamma Ray Spectrometer); HEND (High-Energy Neutron Detector); HiRISE (High Resolution Imaging Science
%Experiment); HRSC (High Resolution Stereo Camera).
%\end{tabnote}
%\end{table}

%  \appendix
%  \include{appendix}
%  \endappendix

% This produces a global reference list for the entire book (don't use)
%\renewcommand{\refname}{Bibliography}% if you prefer this heading
%\bibliography{cplslargesample}\label{refs}
%\bibliographystyle{cambridgeauthordate}
 \printindex

\end{document}